\documentclass[twocolumn,showpacs,preprintnumbers,amsmath,amssymb]{revtex4}
\usepackage{epsfig}

\usepackage[table]{xcolor}
\usepackage{graphicx}
\usepackage{dcolumn}
\usepackage{bm}
\usepackage{esvect}

\usepackage{hyperref}
\hypersetup{colorlinks=true,linkcolor=magenta,anchorcolor=green,citecolor=cyan,filecolor=black,menucolor=black,urlcolor=brown}
\usepackage{slashed}

\newcommand{\be}{\begin{equation}}
\newcommand{\ee}{\end{equation}}
\newcommand{\ba}{\begin{eqnarray}}
\newcommand{\ea}{\end{eqnarray}}

\begin{document}

\title{K-essence scalar dark matter solitons around supermassive black holes}

\author{Philippe Brax}
\affiliation{Universit\'{e} Paris-Saclay, CNRS, CEA, Institut de physique th\'{e}orique, 91191, Gif-sur-Yvette, France}
\author{Jose A. R. Cembranos}
\affiliation{Departamento de  F\'{\i}sica Te\'orica and IPARCOS,\\
Universidad Complutense de Madrid, E-28040 Madrid, Spain}
\author{Patrick Valageas}
\affiliation{Universit\'{e} Paris-Saclay, CNRS, CEA, Institut de physique th\'{e}orique, 91191, Gif-sur-Yvette, France}

\begin{abstract}

We consider scalar dark matter models where the theory has a shift symmetry only broken
by the scalar mass term. We restrict ourselves to K-essence kinetic terms where
the shift symmetric part of the Lagrangian is a function of the first derivatives of the scalar field
only.
When the scalar mass is much larger than the inverse of the astrophysical time and length
scales of interest, these models provide a description of dark matter equivalent to the one
given by theories with only polynomial interactions, in the low-amplitude regime where the
self-interactions are small contributions to the Lagrangian.
In this regime and in the nonrelativistic limit, which apply on large galactic scales, scalar clouds
form solitons with a finite core. This provides an adequate model for dark matter halos with
no singular behavior. Close to the center of galaxies, where a supermassive Black Hole (BH)
resides, we analyze the scalar field distribution and the fate of the dark matter soliton when
subject to the BH gravitational attraction.
We show that the scalar field profile around such a central BH can be described by new
oscillatory solutions of a modified Klein-Gordon equation, which generalize the
harmonic oscillations of free scalar dark matter in a flat environment and the Jacobi elliptic
functions of the $\phi^4$ model. Moreover, we find that, depending on the form of the
K-essence kinetic term, regular solutions can be constructed or not,
which connect the relativistic ingoing wavelike profile of the scalar field at the BH horizon
to the nearly static nonrelativistic soliton at large distance.
These profiles have a constant flux and represent the slow infall of scalar matter into the BH.
We show that this regular behavior is only possible for K-essence functions that satisfy
the usual conditions for the absence of ghosts and gradient instabilities, together with
a new restriction on the growth of the kinetic function $K(X)$ for large argument.
It turns out that the same conditions of stability guarantee that quantum corrections are tamed,
provided that the mass of the scalar field is less than $10^{-3}$ eV and the strong coupling
scale of the model $\Lambda$ is much larger than the scalar mass.

\end{abstract}

\date{\today}

\maketitle


\section{Introduction}
\label{sec:introduction}

Conventional descriptions of dark matter involving heavy particles with weak interactions (WIMPs) have failed to show up so far in all
experiments tracking them from astrophysical scales, i.e. indirect detection, to the laboratory, i.e. direct detection and production at accelerators. Moreover, the confrontation between the predictions of such Cold Dark Matter (CDM) scenarios using  large-scale computer simulations or analytical estimates are in tension with astrophysical observations
\cite{Ostriker:2003qj,Weinberg:2013aya, Pontzen:2014lma}.
The so called `\emph{core-cusp}' \cite{deBlok:2009sp}, `\emph{missing satellites}' \cite{Moore:1999nt} or `\emph{too big to fail}' \cite{BoylanKolchin:2011de} problems are well known examples of these
open questions, that do not have a definitive answer.
This has prompted the search for alternatives to the standard scenario. Axions
\cite{Peccei:1977hh,Wilczek:1977pj,Weinberg:1977ma}
and ALPs (Axion-Like Particles) \cite{Marsh:2015xka}
have been suggested and have particular features which are well-documented. More generally, scalar fields, either fundamental or as an effective low energy description of underlying theories, have been extensively studied in the last decade. Particular emphasis has been put on Fuzzy Dark Matter models (FDM) \cite{Hu:2000ke,Hui:2016ltb}, where a very light  scalar field could form a condensate whose average properties would coincide with the ones expected from dark matter
for the formation of large scale structures
\cite{Johnson:2008se, Hwang:2009js, Park:2012ru, Hlozek:2014lca, Cembranos:2015oya,Cembranos:2016ugq},
but present distinctive features for the behavior at small scales
\cite{Hlozek:2014lca,Schive:2014dra, Broadhurst:2018fei,Ostriker:2003qj,Cembranos:2005us,Chavanis:2011zi,Chavanis:2011zm,Weinberg:2013aya,Pontzen:2014lma,BoylanKolchin:2011de,Moore:1999nt,deBlok:2009sp,Cembranos:2018ulm,Chavanis:2018pkx}.
Indeed, one of the salient features of these models is the wavelike behavior of dark matter on galactic scales, following from the non-negligible role played by the so called
``quantum pressure''. Such models require masses typically less than $10^{-21}$ eV and could be in conflict with a host of astrophysical observations \cite{Armengaud:2017nkf}, such as the spectrum of the Lyman-$\alpha$ forest.

In this paper, we focus on scalar dark matter models where the scalar field has a much larger mass \cite{Brax:2019fzb}. In this regime, the quantum pressure can be neglected on galactic scales and the scalar self-interactions play a dominant role
\cite{Khlopov:1985jw,Goodman:2000tg,Li:2013nal,Suarez:2016eez,Suarez:2015fga,Cembranos:2015oya,Cembranos:2018ulm}.
In particular, they can provide the repulsive pressure which balances the gravitational attraction, allowing for clouds of dark matter to be stable on large scales. Such clouds form soliton-like objects which are candidates for representing dark matter halos with a finite core. This behavior is typically obtained for dark matter scalar fields with
a positive $\phi^4$ self-interaction. Moreover, as shown in \cite{Brax:2019npi}, these solitons are long-lived even when the supermassive BH at the center of the halo is taken into account. Indeed, the lifetime of such objects is longer than the age of the Universe.

Here we consider models of scalar dark matter where the scalar mass term is complemented with K-essence kinetic terms \cite{ArmendarizPicon:2000ah}. On large scales and in the nonrelativistic limit, these models are equivalent to self-interacting models of scalars with polynomial interactions. We extend this analysis to the case where there is a supermassive BH at the center of the galaxy. In this case, the equivalence with polynomial models is more subtle; in particular, we show that regular dark matter profiles with constant scalar fluxes, which must behave as ingoing waves close to the BH horizon, cannot always be connected to the solitonic solution at large radii.
This happens for the $(\partial\phi)^4$ model, where the scalar field cannot sustain a large scalar cloud in the presence of the central BH. We give conditions for the existence of regular solutions where the scalar profile exists and is regular from the BH horizon to spatial infinity. On top of the usual K-essence stability conditions for the absence of ghosts and gradient instabilities, we find that the growth of the K-essence function for large argument cannot be too steep. In this case, this also guarantees that the models are stable under quantum corrections, even though the model becomes nonlinear close to the BH horizon.

The paper is arranged as follows. In section \ref{sec:dark-matter}, we describe the models
of scalar dark matter with nonlinear kinetic terms and connect them in the nonrelativistic regime
with theories that have nonlinear scalar potentials. In section \ref{sec:nonlinear-solution},
we present the nonlinear solutions to the modified Klein-Gordon equation and the constant flux
solutions. In section \ref{sec:small-field-large-radius}, we make the connection between the
nonlinear solutions and the large-radius and nonrelativistic limits.
We also consider the behavior close to the horizon.
In section \ref{sec:quartic-Lagrangian}, we give the example of quartic Lagrangians
for which constant flux solutions connected to stable solitons at large radii do not exist.
We then discuss when global solutions exist in section \ref{sec:power-law}.
Then, in section \ref{sec:explicit}, we give an explicit example of models for which constant
flux solutions up to very large radii exist and  the lifetime of the soliton is larger than the age
of the Universe. In section \ref{sec:renormalization}, we discuss the quantum stability of
these models.
We finally conclude in section \ref{sec:conclusion}.

\section{Dark matter scalar field with derivative self-interactions}
\label{sec:dark-matter}

\subsection{Scalar field action with nonstandard kinetic term}
\label{sec:action}

In this paper, we investigate scenarios where the dark-matter scalar-field action is
\be
S_\phi = \int d^4x \sqrt{-g} \left[ \Lambda^4 K(X) -  \frac{m^2}{2} \phi^2 \right] ,
\label{eq:action-S-def}
\ee
where the normalized kinetic argument $X$ is given by
\be
X = - \frac{1}{2\Lambda^4}   g^{\mu\nu} \partial_\mu \phi \partial_\nu \phi ,
\ee
and we decompose the nonstandard kinetic term $K(X)$ as the sum of the
standard term $X$ and a nonstandard nonlinear contribution $K_{\rm I}$,
\be
K(X) = X + K_{\rm I}(X) .
\ee
We assume that $K_{\rm I}$ admits the small-$X$ expansion
\be
X \ll 1 : \;\;\; K_{\rm I}(X) = \sum_{n\geq 2} \frac{k_n}{n} X^n .
\label{eq:KI-small-X}
\ee
The scale $\Lambda$ plays the role of the strong coupling scale. We will see that the models make sense quantum mechanically even when $X \gg 1$, see section \ref{sec:renormalization}.

As shown in \cite{Brax:2019fzb}, in the nonrelativistic and large-mass regime, where
$K_{\rm I} \ll X$, the small nonlinear correction $K_{\rm I}$ is equivalent
to a small nonlinear potential $V_{\rm I}$, with $V_{\rm I} \ll m^2\phi^2/2$ and
\be
V_{\rm I}(\phi) = \Lambda^4 \sum_{n\geq 4} \frac{\lambda_n}{n} \frac{\phi^n}{\Lambda^n} ,
\label{eq:V-I-K-I}
\ee
with
\be
\lambda_{2n} = - 2 k_n \left( \frac{m^2}{2\Lambda^2} \right)^n .
\ee
This result is obtained at leading order in the large-mass limit, when  the
dynamics are averaged over the fast oscillations $e^{imt}$ driven by the zeroth-order quadratic
Lagrangian $\Lambda^4 X-m^2\phi^2/2$.

In the case of a quartic derivative self-interaction, we obtain
\be
K_{\rm I}(X) = \frac{k_2}{2} X^2 , \;\;\;
V_{\rm I}(\phi) = \frac{\lambda_4}{4} \phi^4 , \;\;\;
\lambda_4 = - k_2 \frac{m^4}{2\Lambda^4} .
\label{eq:lambda4-k2}
\ee
For positive $\lambda_4$, hence negative $k_2$, this gives rise to an effective pressure
on small scales \cite{Brax:2019fzb}.
This leads to a nonzero Jeans length for the growth of cosmological structures
and in virialized halos the scalar field can relax to a static soliton, where the halo
self-gravity is balanced by this effective pressure due to the (derivative) self-interaction.
Therefore, in this paper we focus on the case
\be
\lambda_4 > 0 , \;\;\; k_2 < 0 .
\label{eq:k2-negative}
\ee

\subsection{Isotropic coordinates}
\label{sec:isotropic}

Throughout most of this paper, we work with isotropic coordinates and we consider
static spherically symmetric configurations. Then, the metric can be written as
\be
ds^2 = - f(r) dt^2 + h(r) ( dr^2 + r^2  d\vec\Omega^2 ) .
\label{eq:ds2-Schwarzschild-isotropic}
\ee
We use natural units with $c=1$ throughout this paper.

The spacetime  around the BH can be divided in three regions.
First, from the Schwarzschild radius and up to a radius $r_{\rm NL}$, the metric is in the
strong-gravity regime dominated by the supermassive BH gravity. Then, the metric functions $f(r)$
and $h(r)$ are given by the standard Schwarzschild metric, but written in the isotropic
coordinates $(r,t)$ instead of the usual Schwarzschild coordinates $(\tilde{r},t)$.
This gives \cite{Blau-2017}
\ba
\frac{r_s}{4} < r < r_{\rm NL} : && f(r) = \left( \frac{1-r_s/(4r)}{1+r_s/(4r)} \right)^2 ,
\label{eq:f-def}  \\
&& h(r) = (1+r_s/(4r))^4 .
\label{eq:h-def}
\ea
Here, $r_s=2{\cal G}M_{\rm BH}$ is the Schwarzschild radius of the BH of mass $M_{\rm BH}$,
\be
r_s = 2{\cal G}M_{\rm BH} \simeq \left( \frac{M_{\rm BH}}{10^8 M_\odot} \right)
10^{-8} \,  {\rm kpc} ,
\label{eq:rs-def}
\ee
$r=r_s/4$ is the Schwarzschild radius in the radial isotropic coordinate $r$,
which is related to the usual Schwarzschild radial coordinate $\tilde{r}$ by \cite{Blau-2017}
\be
\tilde{r} > r_s, \;\; r > \frac{r_s}{4} : \;\;\;
\tilde r= r \left(1+\frac{r_s}{4r}\right)^2 .
\label{eq:tilde-r-def}
\ee

Second, beyond $r_{\rm NL}$ and up to $r_{\rm sg}$, the metric is  in the weak-gravity regime while
the gravitational potential remains dominated by the BH itself. This gives
\be
r> r_{\rm NL} : \;\;\; f(r) = 1+2\Phi , \;\; h(r) = 1-2\Phi ,
\label{eq:weak-gravity}
\ee
with
\be
r_{\rm NL} < r \ll r_{\rm sg} : \;\;\; \Phi(r) = -\frac{r_s}{2r}
= -\frac{{\cal G}M_{\rm BH}}{r} .
\label{eq:Phi-BH}
\ee

Third, beyond the radius $r_{\rm sg}$ the metric is also in the weak-gravity regime,
as in Eq.(\ref{eq:weak-gravity}), but the gravitational potential is dominated by the
self-gravity of the dark-matter scalar-field cloud. Then, $\Phi$ is given by the
scalar field Poisson equation
\be
r \gg r_{\rm sg}  : \;\;\; \nabla^2 \Phi = 4\pi{\cal G} \rho_\phi ,
\label{eq:Poisson}
\ee
where $\rho_\phi$ is the scalar field energy density.

\subsection{Equations of motion}
\label{sec:e-o-m}

In the static spherical metric (\ref{eq:ds2-Schwarzschild-isotropic}), the scalar-field
Klein-Gordon equation reads
\be
\frac{\partial}{\partial t} \left[ K' \frac{\partial\phi}{\partial t} \right]
- \sqrt{\frac{f}{h^3}} \frac{1}{r^2}  \frac{\partial}{\partial r} \left[ \sqrt{f h} r^2 K'
\frac{\partial\phi}{\partial r} \right] + f m^2 \phi = 0 ,
\label{eq:KG-phi-1}
\ee
where $K'=dK/dX$ and
\be
X = \frac{1}{2 \Lambda^4 f} \left( \frac{\partial\phi}{\partial t} \right)^2
- \frac{1}{2 \Lambda^4 h} \left( \frac{\partial\phi}{\partial r} \right)^2 .
\label{eq:X-phi}
\ee

\subsection{Large-radius soliton}
\label{sec:large-radius-soliton}

At large radii, $r \gg r_{\rm sg}$, the gravitational field is small and set by the
self-gravity of the scalar cloud.
Therefore, assuming the influence of the BH can indeed be neglected, we recover
the solitonic solution of the dark matter halo as analyzed in \cite{Brax:2019fzb}.
We briefly recall in this section their results, which we will need to set the large-radius
boundary conditions when we analyze the exact solution that takes into account
the central BH.
In this nonrelativistic regime, we can write the real scalar field $\phi$ in terms
of a complex scalar field $\psi$ as
\be
\phi = \frac{1}{\sqrt{2m}} \left( e^{-imt} \psi + e^{imt} \psi^\star \right) .
\label{eq:phi-psi}
\ee
In the large-mass limit, where macroscopic momentum scales are much below $m$,
this actually decomposes $\phi$ in a fast oscillation $e^{\pm imt}$, which is associated
with the zeroth-order of the Klein-Gordon equation (\ref{eq:KG-phi-1}),
$\partial_t^2 \phi + m^2 \phi = 0$, and a slow time and space dependent part
$\psi(r,t)$, which is associated with the variation of gravitational potentials
and matter densities on astrophysical time and length scales.
Next, the dynamics of the complex field $\psi$ can be mapped to an hydrodynamics
problem through the Madelung transformation \cite{Madelung_1927},
\be
\psi = \sqrt{\frac{\rho}{m}} e^{i s} , \;\;\;
\phi = \frac{\sqrt{2\rho}}{m} \cos(m t - s) ,
\label{eq:Madelung}
\ee
where $\rho$ plays the role of the scalar field matter density while the velocity field
${\vec v}$ is defined from the phase $s$ by
\be
\vec{v} = \frac{\vec\nabla s}{m} .
\label{eq:v-s-def}
\ee
Then, the dynamics are governed by the continuity and Euler equations,
\ba
&& \dot\rho + \vec\nabla \cdot ( \rho \vec{v} ) = 0 , \\
&& \dot{\vec v} + ( \vec{v} \cdot \vec\nabla ) \vec v = - \vec\nabla (\Phi+\Phi_{\rm I}) ,
\label{eq:Euler-1}
\ea
where $\Phi$ is the gravitational potential (\ref{eq:Poisson}), where
$\rho_\phi = \rho$, and $\Phi_{\rm I}$ is a repulsive self-interaction potential.
In the quartic case it is given by \cite{Brax:2019fzb}
\be
\Phi_{\rm I}(\rho) = \frac{\rho}{\rho_a} \;\;\; \mbox{with} \;\;\;
\rho_a \equiv \frac{4 m^4}{3\lambda_4} = \frac{8 \Lambda^4}{3 |k_2|} .
\label{eq:Phi-I-def}
\ee
Here we neglected the ``quantum pressure'' $\Phi_{\rm Q}$, associated with the wavelike
nature of the scalar field, because we consider large masses $m \gg 10^{-21} {\rm eV}$,
beyond the ranges associated with Fuzzy Dark Matter scenarios.
The ``pressure'' $\Phi_{\rm I}$ associated with the self-interactions allows the scalar cloud
to reach an hydrostatic equilibrium, where this repulsive self-interaction balances the
self-gravity.
This gives the soliton profile \cite{Brax:2019fzb}
\be
\rho(r) = \rho_{\rm sol}(0) \frac{\sin(r/r_a)}{r/r_a} , \;\;\;
\Phi_{\rm I}(r) = \Phi_{\rm I, sol}(0) \frac{\sin(r/r_a)}{r/r_a} ,
\label{eq:soliton-rho-PhiI}
\ee
with ${\vec v} = 0$ and
\be
r_a = \frac{1}{\sqrt{4\pi{\cal G} \rho_a}} = \sqrt{\frac{3\lambda_4}{2}}
\frac{M_{\rm Pl}}{m^2} ,
\label{eq:r_a-rho_a}
\ee
where we introduced the reduced Planck mass $M^2_{\rm Pl}= 1/(8\pi{\cal G})$.
The soliton has a flat inner core and a finite radius $R_{\rm sol}=\pi r_a$, which can reach
galactic size depending on the value of $\lambda_4$. More precisely, we can also
write (\ref{eq:r_a-rho_a}) as
\be
\lambda_4 = \left( \frac{r_a}{20 \, {\rm kpc}} \right)^2
\left( \frac{m}{1 \, {\rm eV}} \right)^4 .
\ee
The constraint that the scalar field behaves as pressureless dark matter at the
background level up to the radiation-matter equality, at redshift $z_{\rm eq}$,
implies \cite{Brax:2019fzb} $\lambda_4 \lesssim (m/1 {\rm eV})^4$, therefore we actually
have $r_a \lesssim 20 \, {\rm kpc}$.

Inside the soliton, the hydrostatic equilibrium condition in Eq.(\ref{eq:Euler-1})
gives $\vec\nabla (\Phi+\Phi_{\rm I}) =0$, and we have
\be
r \leq R_{\rm sol} : \;\;\; \Phi+\Phi_{\rm I} = \alpha ,
\label{eq:alpha-def}
\ee
where $\alpha$ is a constant, given by the value of the Newtonian potential
at the boundary of the soliton,
\be
\alpha = \Phi(R_{\rm sol}) ,
\label{eq:alpha-Phi-Rs}
\ee
as $\Phi_{\rm I}(R_{\rm sol})=0$.
In terms of the scalar fields $\psi$ and $\phi$ this gives \cite{Brax:2019fzb}
\be
\psi = \sqrt{\frac{\rho}{m}} e^{-i\alpha m t} , \;\;\; \mbox{hence} \;\;\;s= - \alpha m t,
\ee
and
\be
\phi = \frac{\sqrt{2\rho}}{m} \cos[ (1+\alpha) m t ] .
\label{eq:phi-soliton}
\ee

\section{Nonlinear global solution}
\label{sec:nonlinear-solution}

\subsection{Oscillating solution in the large-mass limit}
\label{sec:large-mass-solution}

As in \cite{Brax:2019npi}, where we considered the case of a scalar field with a standard
kinetic term and a self-interaction potential, we look for a solution in the large-mass
limit.
Then, the field oscillates with a very high frequency determined by $m$,
if we only keep the zeroth-order terms that give the standard Klein-Gordon equation
$\partial_t^2 \phi + m^2 \phi = 0$.
However, the nonlinearity associated with the higher-order kinetic factor $K_{\rm I}$
transforms this harmonic oscillator into an anharmonic oscillator, with parameters
that slowly change with radius as dictated by the radial derivative term.
In a fashion similar to the case of the quartic potential studied in \cite{Brax:2019npi},
we look for a solution of the nonlinear Klein-Gordon equation (\ref{eq:KG-phi-1})
of the form
\be
\phi(r,t) = \phi_0(r) \, {\rm ck}[ \omega(r) t - {\bf Q}(r) \beta(r), \mu(r) ] .
\label{eq:phi-ck-def}
\ee
Here ${\rm ck}(u,\mu)$ is the extension of the harmonic cosine $\cos(u)$, obtained
for the free massive scalar field, and of the Jacobi elliptic function ${\rm cn}(u,k)$,
obtained for the quartic potential \cite{Brax:2019npi}, to the case of derivative
self-interactions (the letter ``k'' refers to the ``kinetic'' nonlinearity).
For $\mu=0$ we will recover the harmonic cosine, ${\rm ck}(u,0)=\cos(u)$, and
for nonzero $\mu$ we will have an anharmonic oscillator, associated with the kinetic factor
$K_{\rm I}$ that adds nonlinear contributions to the Klein-Gordon equation.
The factor ${\bf Q}(r)$ is defined as ${\bf Q}(r) \equiv {\bf Q}[\mu(r)]$, where
${\bf Q}(\mu)$ is the quarter of the period of the oscillator ${\rm ck}(u)$ for parameter
$\mu$. It is introduced in (\ref{eq:phi-ck-def}) for future convenience,
to simplify Eq.(\ref{eq:dphi-dr-ck}) below.
Thus, $\mu$ and ${\bf Q}$ play the role of the modulus $k$ and the complete
elliptic integral ${\bf K}$ that appears in the case of the quartic potential
\cite{Brax:2019npi}.
At this stage, ${\rm ck}(u,\mu)$ is not defined yet and it will be determined below from
the analysis of the nonlinear Klein-Gordon equation.

The expression (\ref{eq:phi-ck-def}) is understood as the leading-order approximation
in the limit $m \to \infty$, where spatial gradients of the functions
$\phi_0,\omega,{\bf Q},\beta$ and $\mu$ are much below $m$ (i.e. $\partial_r \ll m$),
whereas both $\omega$ and $\beta$ are of order $m$.
Thus, the scalar field shows fast oscillations with time at each radius,
at a frequency and a phase of order $m$, with a slow modulation in space
of the oscillation characteristics. This behavior relies on the large separation
of scales $\partial_r \ll m$, which in our case corresponds to $r_s \gg m$,
as radial derivatives typically scale as $\partial_r \sim 1/r \lesssim 1/r_s$
beyond the horizon.

To ensure that spatial gradients do not increase with time, the scalar field must
oscillate with the same frequency over all radii, with a common period $T=2\pi/\omega_0$,
where $\omega_0$ is the common angular frequency.
Otherwise, there would be a secular growth with time of the phase difference between
neighbouring points, hence a secular growth of radial gradients.
Since the period of the function ${\rm ck}(u)$ for parameter $\mu$ is $4{\bf Q}$,
this implies $\omega T=4{\bf Q}$ and $\omega(r)$ is fully determined
by the oscillatory parameter $\mu(r)$ as
\be
\omega(r) = \frac{2{\bf Q}(r)}{\pi} \omega_0 .
\label{eq:omega-omega0}
\ee
As we shall check in section~\ref{sec:soliton-boundary} and Eq.(\ref{eq:omega0-Phi-k2})
below, the common frequency $\omega_0$ must match the oscillation found at large radii
in the soliton solution (\ref{eq:phi-soliton}). This implies
\be
\omega_0 = (1+\alpha) m .
\label{eq:omega0-alpha}
\ee
As we shall see in section~\ref{sec:nonlinear-oscillator}, the oscillating function obeys
the Fourier series (\ref{eq:ck-Fourier}) below. Substituting into Eq.(\ref{eq:phi-ck-def}) gives
\be
\phi = \phi_0(r) \sum_{n=0}^{\infty} a_{2n+1}(r)
\cos [ (2n+1) (\omega_0 t - \pi \beta(r)/2) ] ,
\label{eq:cn-series}
\ee
with $a_{2n+1}(r) \equiv a_{2n+1}[\mu(r)]$.
Thanks to the relation (\ref{eq:omega-omega0}), we can see that the scalar field shows
a coherent nonlinear oscillation over all radii, at the common angular frequency $\omega_0$.

From Eq.(\ref{eq:phi-ck-def}), the time derivative of the scalar field is
\be
\frac{\partial\phi}{\partial t} = \phi_0 \omega \frac{\partial {\rm ck}}{\partial u} .
\label{eq:dphi-dt-ck}
\ee
At leading order in the large-$m$ limit, the radial derivative reads from
Eq.(\ref{eq:cn-series}) as
\be
\frac{\partial\phi}{\partial r} \simeq  \phi_0 \sum_{n=0}^{\infty} a_{2n+1}
(2n+1) \frac{\pi \beta'}{2}
\sin \left[ (2n+1) \left( \omega_0 t - \frac{\pi \beta}{2} \right) \right] ,
\label{eq:phi-Fourier-series}
\ee
where $\beta' = d\beta/dr$.
Here we only kept the term of order $m$, as we assume that $\phi_0$, $\mu$
and $\beta$ are slow functions of $r$, but $\beta$ is of order $m$.
Thus, the factor $\beta'$ yields an additional power of $m$ as compared with
$\phi_0'$ or $a_{2n+1}'$.
Comparing with the Fourier series of $\frac{\partial{\rm ck}}{\partial u}$, obtained
from Eq.(\ref{eq:ck-Fourier}) below, this gives
\be
\frac{\partial\phi}{\partial r} \simeq - \phi_0 {\bf Q} \beta'
\frac{\partial {\rm ck}}{\partial u} ,
\label{eq:dphi-dr-ck}
\ee
in this large-$m$ limit.
The factor ${\bf Q}$ was introduced in Eq.(\ref{eq:phi-ck-def}) to simplify
this radial derivative (the change $\beta \to \beta/{\bf Q}$ would change
the factor ${\bf Q}\beta'$ above to $\beta'-\beta {\bf Q}'/{\bf Q}$).
Indeed, if we had written $\phi = \phi_0 \, {\rm ck}[\omega t - \beta,\mu]$,
we would have found that a slow radial change of $\mu$, hence of the period $4{\bf Q}$,
generates a leading-order change of the phase of the oscillation and must be taken into
account. This effect is automatically taken care of by renormalizing the phase $\beta$
by the quarter of period ${\bf Q}$ in Eq.(\ref{eq:phi-ck-def}).

In this approximation, the kinetic term $X$ of Eq.(\ref{eq:X-phi}) reads
\be
X = \frac{\phi_0^2}{2 \Lambda^4 f} \left[ \omega^2 - \frac{f}{h} ({\bf Q} \beta')^2 \right]
\left( \frac{\partial {\rm ck}}{\partial u} \right)^2 ,
\label{eq:X-ck2}
\ee
and the nonlinear Klein-Gordon equation (\ref{eq:KG-phi-1}) becomes
\be
\left[ \omega^2 - \frac{f}{h} ( {\bf Q} \beta' )^2 \right]
( 1 + \tilde{K}_{\rm I} )  \frac{\partial^2 {\rm ck}}{\partial u^2} + f m^2 {\rm ck} = 0 ,
\label{eq:KG-tK-I}
\ee
where we defined
\be
\tilde{K}_{\rm I}(X) \equiv K'_{\rm I} + 2 X K''_{\rm I} ,
\label{eq:tK-I-def}
\ee
where the prime denotes the derivative with respect to $X$.
If the self-interaction term $\tilde{K}_{\rm I}$ vanishes we recover the harmonic
oscillator. For nonzero self-interaction, we obtain an anharmonic oscillator,
with a derivative nonlinearity. The kinetic argument $X$ of
Eq.(\ref{eq:X-ck2}) can be decomposed in a time-independent prefactor,
with a slow radial dependence, and a fast oscillatory term. Thus, we define
the prefactor $\mu(r)$ by
\be
\mu(r) \equiv \frac{\phi_0^2}{2 \Lambda^4 f} \left[ \omega^2 - \frac{f}{h}
({\bf Q} \beta')^2 \right] ,
\label{eq:mu-r-def}
\ee
so that we have
\be
X = \mu(r) \left( \frac{\partial {\rm ck}}{\partial u} \right)^2 .
\label{eq:X-mu-ck2}
\ee

Now, let us define an oscillatory function ${\rm ck}(u,\mu)$, of argument $u$ and
parameter $\mu$, by the differential equation
\be
\frac{\partial^2 {\rm ck}}{\partial u^2} + {\rm ck}
+ \tilde{K}_{\rm I}\left[ \mu \left( \frac{\partial {\rm ck}}{\partial u} \right)^2 \right]
\frac{\partial^2 {\rm ck}}{\partial u^2} \equiv 0 ,
\label{eq:ck-ODE-def}
\ee
and the initial conditions
\be
{\rm ck}(0,\mu) \equiv 1 , \;\;\;
\frac{\partial {\rm ck}}{\partial u}(0,\mu) \equiv 0 .
\label{eq:ck-init}
\ee
From Eq.(\ref{eq:KI-small-X}) we have $\tilde{K}_{\rm I} \to 0$ for $X \to 0$.
Therefore, in the limit $\mu\to 0$ the nonlinear differential equation
(\ref{eq:ck-ODE-def}) simplifies to the linear harmonic oscillator,
$\frac{\partial^2 {\rm ck}}{\partial u^2} + {\rm ck} = 0$,
and with the initial conditions (\ref{eq:ck-init}) we recover the cosine function,
\be
{\rm ck}(u,0) = \cos(u) .
\label{eq:ck-mu0-cos}
\ee
The initial conditions (\ref{eq:ck-init}) do not entail any loss of generality.
They mean that ${\rm ck}(u)$ oscillates over the range
\be
-1 \leq {\rm ck} \leq 1 ,
\ee
and it starts at a maximum at $u=0$. The normalization to unity of the amplitude
of ${\rm ck}$ simply sets the normalization of the amplitude $\phi_0$ in
Eq.(\ref{eq:phi-ck-def}), while the choice $u=0$ for a maximum sets an integration
constant for the phase $\beta$ or the origin of time $t$.
Also, the choice of unity for the first two coefficients in the differential equation
(\ref{eq:ck-ODE-def}) does not lead to a loss of generality. It sets the normalization
of the argument $u$, that is, the period $4{\bf Q}$ of the oscillator.

Then, comparing the definition (\ref{eq:ck-ODE-def}) with the nonlinear Klein-Gordon
equation (\ref{eq:KG-tK-I}), we can see that this equation of motion is satisfied
if ${\rm ck}(u,\mu)$ is the function defined in Eq.(\ref{eq:ck-ODE-def}), provided
we have
\be
f m^2 = \omega^2 - \frac{f}{h} ( {\bf Q} \beta' )^2 ,
\label{eq:omega2-betap}
\ee
and the parameter $\mu$ of the oscillator (\ref{eq:ck-ODE-def}) is set to the
value $\mu(r)$ of Eq.(\ref{eq:mu-r-def}).
Combining with Eqs.(\ref{eq:omega-omega0}) and (\ref{eq:mu-r-def}), we obtain
\ba
&& \frac{\pi^2 f}{4 h} \beta'^2 = \omega_0^2 - \frac{\pi^2 m^2 f}{4 {\bf Q}^2} ,
\label{eq:beta-1}
\\
&& \phi_0^2 = \frac{2\Lambda^4}{m^2} \mu .
\label{eq:phi0-mu}
\ea
These two equations take the same form as Eqs.(62)-(63) obtained in \cite{Brax:2019npi}
for the case of a quartic potential.
For a given radial function $\mu(r)$, they provide the phase $\beta(r)$ and the amplitude
$\phi_0(r)$. This fully determines the oscillating solution (\ref{eq:phi-ck-def}),
as the frequency $\omega(r)$ is given by Eq.(\ref{eq:omega-omega0}) and ${\bf Q}(r)$
is determined by $\mu(r)$ as the quarter of period of the oscillator (\ref{eq:ck-ODE-def}).
Equation (\ref{eq:phi0-mu}) provides at once the constraint
\be
\mu \geq 0 .
\label{eq-mu-positive}
\ee

We can see in Eq.(\ref{eq:phi0-mu}) that for low scalar-field amplitudes,
$\phi_0 \to 0$, we recover the harmonic oscillator as $\mu \to 0$.
This corresponds to the nonrelativistic and small-field limit, found for instance
for the soliton solution (\ref{eq:phi-soliton}), where the higher-order contributions
to the scalar field Lagrangian are small, $K_{\rm I} \ll X$, and the Klein-Gordon
equation reduces to the harmonic oscillator at leading order.

Equation (\ref{eq:beta-1}) appears as  the generalization of the Euler equation at leading order,
$\pi \beta'/(2m)$ playing the role of the radial velocity $v_r=m^{-1} ds/dr$ and
$\pi \beta/2$  the role of the phase $s$.
Using Eqs.(\ref{eq:omega0-alpha}) and (\ref{eq:beta-1}) we can write
\be
v_r \equiv \frac{\pi \beta'}{2m} = - \sqrt{ \frac{h}{f} }
\sqrt{ (1+\alpha)^2 - \frac{\pi^2 f}{4 {\bf Q}^2} } .
\label{eq:vr-sqrt}
\ee
We refine this analogy below.

\subsection{Nonlinear oscillator}
\label{sec:nonlinear-oscillator}

We now investigate the behavior of the anharmonic oscillator (\ref{eq:ck-ODE-def}).
This equation of motion can be integrated once, after multiplying
by $2 \frac{\partial{\rm ck}}{\partial u}$.
With the initial conditions (\ref{eq:ck-init}), this gives
\be
\left( \frac{\partial{\rm ck}}{\partial u} \right)^2 + {\rm ck}^2
+ \frac{1}{\mu} G_{\rm I} \left[ \mu \left( \frac{\partial{\rm ck}}{\partial u} \right)^2
\right] = 1 ,
\label{eq:ck-E}
\ee
where we introduced the function $G_{\rm I}(X)$ defined by
\be
G_{\rm I}(0) \equiv 0 , \;\;\;
G_{\rm I}' \equiv \tilde{K}_{\rm I} = K'_{\rm I} + 2 X K''_{\rm I} ,
\label{eq:GI-def}
\ee
hence
\be
G_{\rm I}(X) = 2 X K'_{\rm I}(X) - K_{\rm I}(X) .
\label{eq:GI-KI-def}
\ee
This corresponds to the conservation of energy of the nonlinear oscillator,
which oscillates over the range $-1 \leq {\rm ck}(u) \leq 1$ with a period that we
denote by $4 {\bf Q}(\mu)$.
For small $\mu$ the term $G_{\rm I}$ is a small correction and ${\rm ck}(u)$ closely
follows $\cos(u)$. For larger $\mu$, the higher-order contribution $G_{\rm I}$ becomes
important and the oscillations are more strongly deformed. Depending on the function
$G_{\rm I}$ the periodic oscillatory behavior may eventually disappear.
Introducing the function $G(X)$ by
\be
G(X) \equiv X+G_{\rm I}(X) = 2 X K'(X) - K(X) ,
\label{eq:G-K-def}
\ee
the conservation equation (\ref{eq:ck-E}) can be inverted as
\be
\left( \frac{\partial{\rm ck}}{\partial u} \right)^2 = \frac{1}{\mu}
G^{-1}\left[ \mu (1-{\rm ck}^2) \right] ,
\label{eq:dck-du-Gm1}
\ee
where $G^{-1}$ is the inverse function of $G$, $G[G^{-1}(y)]=y$.
For small $X$ we have $G(X) \simeq X$ and $G^{-1}(y) \simeq y$.
Thus, near the maximum ${\rm ck}(u=0)=1$ we have
$(\frac{\partial{\rm ck}}{\partial u})^2 \simeq 1-{\rm ck}^2$.
To the right of the maximum, $u \gtrsim 0$, the function ${\rm ck}$
decreases below unity with a negative slope given by
$\frac{\partial{\rm ck}}{\partial u} \simeq - \sqrt{1-{\rm ck}^2}$.
We can build the periodic function ${\rm ck}(u)$ from its first quarter of period,
where $0 \leq u \leq {\bf Q}$ and $1 \geq {\rm ck} \geq 0$,
if we can solve the equation (\ref{eq:dck-du-Gm1}) until the point ${\rm ck}=0$
in a finite time $u={\bf Q}$.
For a given parameter $\mu \geq 0$, this requires that $G^{-1}(y)$ is
well defined and positive over $0 \leq y \leq \mu$.
Since $G^{-1}$ is defined from $G(X)$ with $X \geq 0$, starting from $G \simeq X$ at low
$X$, we can see that $G^{-1}$ is positive from the parametric representation
$\{y,G^{-1}\} = \{G,X\}$. Moreover, it is well defined up to $\mu$ if
$G(X)$ is monotonically increasing over $0 \leq X \leq X_\mu$,
where $X_\mu$ is defined by $G(X_\mu) \equiv \mu$.
From Eq.(\ref{eq:G-K-def}) this implies
\be
0 \leq X \leq X_\mu : \;\;\; G'(X) = K' + 2 X K'' > 0 .
\label{eq:Gp-positive}
\ee
Then, the dynamics can be solved by quadrature,
\be
0 \leq {\rm ck} \leq 1: \;\;\; u = \int_{\rm ck}^1 d\psi \sqrt{ \frac{\mu}
{G^{-1}[\mu (1-\psi^2)]} } .
\ee
For $\mu \to 0$ we recover the arccosine function.
Then, the quarter ${\bf Q}$ of the period is given by
\be
{\bf Q}(\mu) = \int_0^1 d\psi \sqrt{ \frac{\mu}{G^{-1}[\mu (1-\psi^2)]} } .
\label{eq:Q-mu}
\ee
Therefore, we can build an oscillatory solution (\ref{eq:phi-ck-def}) for the scalar
field $\phi(r,t)$ if for all values $\mu(r)$ that are reached beyond the horizon
the function ${\bf Q}(\mu)$ defined by Eq.(\ref{eq:Q-mu}) is well defined and finite.
As for the cosine, we can build ${\rm ck}(u)$ over all real $u$ from the first
quarter of period, $0 \leq u \leq {\bf Q}$. We first extend up to the
first minimum, ${\bf Q} \leq u \leq 2 {\bf Q}$, with ${\rm ck}(2{\bf Q}-u)=-{\rm ck}(u)$.
Second, we extend to the first minimum on the left, at $u=-2{\bf Q}$, with
${\rm ck}(-u)= {\rm ck}(u)$. Third, we extend from $-2{\bf Q} \leq u \leq 2{\bf Q}$
to all real $u$ with the periodicity ${\rm ck}(u+4{\bf Q})= {\rm ck}(u)$.
In other words, like the cosine, the periodic function ${\rm ck}(u)$ is even,
of period $4{\bf Q}$, and verifies ${\rm ck}(2{\bf Q}-u)=-{\rm ck}(u)$,
\ba
&& {\rm ck}(-u)= {\rm ck}(u) , \;\;\; {\rm ck}(u+4{\bf Q})= {\rm ck}(u) , \nonumber \\
&& {\rm ck}(2{\bf Q}-u)= - {\rm ck}(u) .
\ea
This implies that its Fourier series takes the form
\be
{\rm ck}(u,\mu) = \sum_{n=0}^{\infty} a_{2n+1}(\mu) \, \cos \left[(2n+1) u
\frac{2\pi}{4{\bf Q}} \right] .
\label{eq:ck-Fourier}
\ee

From Eq.(\ref{eq:KI-small-X}) we obtain the series expansions
\ba
&& X \to 0 : \;\;\; G(X) = X + \frac{3}{2} k_2 X^2 + \dots , \\
&& y \to 0 : \;\;\; G^{-1}(y) = y - \frac{3}{2} k_2 y^2 + \dots ,
\ea
and substituting into Eq.(\ref{eq:Q-mu}) yields
\be
{\bf Q}(\mu) = \frac{\pi}{2} \left( 1 + \frac{3 k_2}{8} \mu + \dots \right) .
\label{eq:Q-mu-series}
\ee

\subsection{Steady state and constant flux}
\label{sec:steady-quartic}

As we have seen above, Eqs.(\ref{eq:beta-1})-(\ref{eq:phi0-mu}) determine the
scalar field solution (\ref{eq:phi-ck-def}) as a function of $\mu(r)$,
but we have not specified the radial profile of $\mu(r)$ yet.
This will be provided by the condition of constant radial flux, after averaging
over the fast oscillation of angular frequency $\omega_0$.
The relativistic counterpart of the continuity equation is the component
$\nu=0$ of the conservation equations $\nabla_\mu T^\mu_\nu = 0$.
The energy-momentum tensor of the scalar field $\phi$ with the action
(\ref{eq:action-S-def}) reads
\be
\rho_\phi \equiv - T^0_0 = \frac{K'}{f} \left( \frac{\partial\phi}{\partial t} \right)^2
- \Lambda^4 K + \frac{m^2}{2} \phi^2 ,
\ee
and
\be
T^r_0 = \frac{K'}{h} \frac{\partial\phi}{\partial r} \frac{\partial\phi}{\partial t} .
\ee
The conservation equation $\nabla_\mu T^\mu_0 = 0$ becomes
\be
\frac{\partial \rho_\phi}{\partial t} - \frac{1}{\sqrt{f h^3} r^2}
\frac{\partial}{\partial r} \left[ \sqrt{f h^3} r^2 T^r_0 \right] = 0 .
\label{eq:continuity-rho-Tr0}
\ee
At  leading order, this continuity equation is satisfied by the balance between
fast oscillatory terms. However, to ensure that secular terms do not appear at
subleading order, we clearly require that in the steady state the averaged value
of $\rho_\phi$ over one oscillation period should not depend on time.
This gives the condition of constant flux $F$,
\be
F = - \sqrt{f h^3} r^2 \langle T^r_0 \rangle = \sqrt{f h} r^2 \phi_0^2 \omega {\bf Q} \beta'
\langle K' \left( \frac{\partial {\rm ck}}{\partial u} \right)^2 \rangle ,
\label{eq:F-Tr0-betap}
\ee
where $\langle \dots \rangle$ denotes the average over one oscillation period
$T=2\pi/\omega_0$ and we used Eqs.(\ref{eq:dphi-dt-ck}) and (\ref{eq:dphi-dr-ck}).

Using Eqs.(\ref{eq:omega-omega0}), (\ref{eq:omega0-alpha}), (\ref{eq:phi0-mu})
and (\ref{eq:vr-sqrt}), we can write the flux in terms of $\mu(r)$,
\be
F = F_s x^2 h \left( \frac{2 {\bf Q}}{\pi} \right)^2 C_\mu
\mu \sqrt{ 1 - \frac{\pi^2 f}{(1+\alpha)^2 4 {\bf Q}^2} }  \, ,
\label{eq:Flux-Fs}
\ee
where we defined the dimensionless radial coordinate
\be
x = \frac{r}{r_s} \geq \frac{1}{4} ,
\ee
the characteristic flux
\be
F_s = - r_s^2 2 \Lambda^4 (1+\alpha)^2 ,
\label{eq:Fs-def}
\ee
and the average
\be
C_\mu = \langle K' \left( \frac{\partial {\rm ck}}{\partial u} \right)^2 \rangle  .
\label{eq:C-mu-def}
\ee
Using Eq.(\ref{eq:dck-du-Gm1}) we obtain
\ba
C_\mu &= & \frac{1}{{\bf Q}} \int_0^{{\bf Q}} du \, K'
\left( \frac{\partial {\rm ck}}{\partial u} \right)^2    \\
& = & \frac{1}{{\bf Q} \sqrt{\mu}} \int_0^1 d\psi \, \sqrt{ G^{-1}[\mu (1-\psi^2)] }
\nonumber \\
&& \times K' \left[ G^{-1}[\mu (1-\psi^2)] \right] .
\label{eq:Cmu}
\ea
Then, for a given value of the flux $F$, the profile $\mu(r)$ of this steady state
solution is determined by Eq.(\ref{eq:Flux-Fs}). The parameter $\mu$ at each radius,
$x=r/r_s$, is such that the right-hand side in Eq.(\ref{eq:Flux-Fs}) is equal to $F$.

From Eq.(\ref{eq:Q-mu-series}), we can see that for $k_2<0$ the period $4{\bf Q}$
decreases for higher $\mu$ (at least until higher-order terms become relevant).
On the other hand, for the square root in Eq.(\ref{eq:Flux-Fs}) to be well defined,
its argument must be positive, which implies ${\bf Q}^2 > \pi^2 f/[4(1+\alpha)^2]$.
Therefore, ${\bf Q}$ cannot be too small, which typically means that $\mu$ cannot be
too large and must obey the upper bound $\mu_+$,
\be
0 \leq \mu \leq \mu_+(x)  \;\;\; \mbox{with} \;\;\;
{\bf Q}^2(\mu_+) = \frac{\pi^2 f}{4(1+\alpha)^2} .
\label{eq:mu+-def}
\ee
Depending on the function ${\bf Q}(\mu)$ and the radius $r$, this upper bound may be finite
or pushed to infinity.
If it is finite, the flux $F(\mu,x)$ as a function of $\mu$ for fixed $x$ vanishes
at both ends, $\mu=0$ and $\mu=\mu_+(x)$. Typically, as in the case of the quartic potential
studied in \cite{Brax:2019npi}, $F(\mu,x)$ shows a single peak between these two endpoints,
so that at each radius there are two solutions $\mu_1<\mu_2$ to Eq.(\ref{eq:Flux-Fs}),
associated with the crossing of the value $F$ along the left and right sides of the peak.

These two solutions are associated with low-velocity and high-velocity
branches. The boundary condition at large radii associated with the soliton
described in section~\ref{sec:large-radius-soliton} selects the low-velocity branch
(as the soliton corresponds to a static regime), while the boundary condition at the
Schwarzschild radius selects the high-velocity branch (as near the horizon the
self-interactions can no longer resist the BH gravity and the fluid falls inward
with a relativistic velocity).
Then, there is a unique critical value $F_c$ for the flux, such that these two branches
connect at some intermediate radius $x_\star$ and the solution can smoothly switch from
the low-velocity branch at large radii to the high-velocity branch at small radii.
This gives rise to a picture that is similar to the unique transonic solution found in the
case of the hydrodynamic infall of relativistic fluids into a BH \cite{Michel:1972}.

\section{Connecting the small-field large-radii regime to the Black Hole horizon}
\label{sec:small-field-large-radius}

\subsection{Small-$\mu$ regime}
\label{sec:small-mu}

At large radii, the metric function $f(r)$ is close to unity, up to deviations of order
$10^{-5}$ as in Eq.(\ref{eq:weak-gravity}). The constant $\alpha$ is also of order
$10^{-5}$ from Eq.(\ref{eq:alpha-Phi-Rs}), as this is the typical magnitude of the
gravitational potential in galactic halos.
Then, from the expansion (\ref{eq:Q-mu-series}) we can see that the bound
(\ref{eq:mu+-def}) is reached for a small value of $\mu$, typically
$|k_2| \mu_+ \sim 10^{-5}$.
Therefore, we investigate in this section the regime $\mu \ll 1$, where the function
${\rm ck}(u)$ is close to $\cos(u)$, as seen in (\ref{eq:ck-mu0-cos}),
and ${\bf Q}\simeq \pi/2$, $C_\mu \simeq 1/2$.
Therefore, at large radii the global solution (\ref{eq:phi-ck-def}) becomes
\be
r \gg r_s : \;\;\; \phi \simeq \phi_0(r) \cos[ \omega_0 t - \pi \beta(r)/2]
\label{eq:phi-small-mu}
\ee
and $\mu \ll 1$, assuming that $k_2$ is of order unity.

At lowest order over $\mu$, $\Phi$ and $\alpha$, the flux (\ref{eq:Flux-Fs}) reads
\be
\{ \mu, \Phi, \alpha \} \ll 1 : \;\;\;
\frac{F}{F_s} \simeq \frac{x^2}{2} \mu \sqrt{2\alpha-2\Phi+ 3 k_2\mu/4} ,
\label{eq:Flux-small-mu}
\ee
where we used Eqs.(\ref{eq:weak-gravity}) and (\ref{eq:Q-mu-series}).
Then, the upper bound $\mu_+$ introduced in (\ref{eq:mu+-def}), associated
with the zero of the square root, is given by (we recall that $k_2 <0$)
\be
r \gg r_s : \;\;\; \mu_+(r) = \frac{8 (\alpha-\Phi)}{3 |k_2|}
= \frac{4 (\alpha-\Phi) m^4}{3 \lambda_4 \Lambda^4} .
\ee
This explicitly shows that $|k_2|\mu \ll 1$ at large radii.
From Eq.(\ref{eq:X-mu-ck2}) we also have $X \lesssim \mu$ and
$\langle X \rangle \simeq \mu/2$.
Then, the nonlinear contribution $K_{\rm I}$ to the kinetic term in the
scalar field Lagrangian is much smaller than the standard linear term $X$.
This is the nonrelativistic regime, which can be described by the complex scalar
field $\psi$ or the hydrodynamical picture $\{\rho,\vec{v}\}$ as in (\ref{eq:Madelung}).
This covers the soliton solution (\ref{eq:soliton-rho-PhiI}), at large radii
$r \gg r_{\rm sg}$ dominated by the scalar cloud self-gravity, as well as the
smaller radii $r_{\rm NL} \ll r < r_{\rm sg}$ dominated by the BH gravity in the weak
field regime.

The flux (\ref{eq:Flux-small-mu}) as a function of $\mu$, for a given radius $x$,
vanishes at both ends of the allowed range $0\leq \mu \leq \mu_+(x)$ and shows
a single peak at the position $\mu_{\rm peak}$ and height $F_{\rm peak}$, with
\be
\mu_{\rm peak} = \frac{2}{3} \mu_+ , \;\;\;\;\;
\frac{F_{\rm peak}}{F_s} = \frac{4 x^2}{3 |k_2|}
\left[ \frac{2(\alpha-\Phi)}{3} \right]^{3/2} .
\label{eq:Fm-small-mu}
\ee
In the weak gravity regime (\ref{eq:Phi-BH}) dominated by the BH, we have
$|\Phi| \gg |\alpha|$ and $\Phi \propto x^{-1}$.
Therefore, in the range $r_{\rm NL} \ll r \ll r_{\rm sg}$ the peak height $F_{\rm peak}$
grows with the radius as $F_{\rm peak} \propto x^{1/2}$.
At larger radii, $r_{\rm sg} \ll r \ll R_{\rm sol}$, dominated by the scalar cloud gravity,
$\Phi$ is almost constant and set by its value inside the soliton core,
so that $F_{\rm peak}$ grows with radius as $F_{\rm peak} \propto x^2$.
Thus, in agreement with Figs.~\ref{fig_flux-quartic} and \ref{fig_flux}
below, we find at large radii a universal
behavior set by the weak-field and nonrelativistic regime and the first higher-order
contribution $k_2 X^2/2$ to the Lagrangian kinetic term.
The flux $F(\mu,x)$ shows a single peak with a height that grows with the distance
from the central BH.

As recalled above, this means that for a given value of the flux $F>0$, which defines
the steady state solution, the equation (\ref{eq:Flux-small-mu}) has two solutions
$\mu_1(x)$ and $\mu_2(x)$ at each radius $x$, with
$0 < \mu_1 < \mu_{\rm peak} < \mu_2 < \mu_+$, provided that $F<F_{\rm peak}$.
They correspond to the crossing of the value $F$ along the two sides of the peak.
Since the peak height $F_{\rm peak}$ grows with the radius, these solutions are increasingly
close to the endpoints $0$ and $\mu_+$. In other words, as the factor $x^2$
becomes large in Eq.(\ref{eq:Flux-small-mu}), either $\mu$ or
$\sqrt{2\alpha-2\Phi+ 3 k_2\mu/4}$ goes to zero as $1/x^2$.
This gives the two asymptotic solutions $\mu_1<\mu_2\ll 1$,
\ba
&& x \gg 1 : \;\;\; \mu_1(x) = \frac{\sqrt{2}F}{F_s \sqrt{\alpha-\Phi} x^2} + \dots ,
\label{eq:mu1-def} \\
&& \; \mu_2(x) \simeq \frac{8 (\alpha-\Phi)}{3|k_2|} - \frac{3|k_2|}{4}
\left( \frac{F}{F_s (\alpha-\Phi) x^2} \right)^2
+ \dots \hspace{0.7cm}
\label{eq:mu2-def}
\ea
where the dots stand for higher-order terms over $1/x^2$.

At lowest order in $\mu$, $\Phi$ and $\alpha$, the velocity $v_r$ of Eq.(\ref{eq:vr-sqrt})
reads
\be
x \gg 1 : \;\;\; v_r \simeq - \sqrt{2 \alpha- 2 \Phi+ 3 k_2 \mu/4} .
\ee
As $\mu_1 \ll \mu_2$, we can see that the left and right branches leads to the different
behaviors
\ba
x \gg 1 : && v_{r_1} \simeq - \sqrt{2 (\alpha-\Phi) } , \label{eq:vr-1} \\
&& v_{r_2} \simeq - \frac{3|k_2| F}{4 F_s (\alpha-\Phi) x^2} \to 0 .
\label{eq:vr-2}
\ea
Thus, $\mu_1$ is the high-velocity branch, as $v_{r_1}$ is of the order
of the free-fall velocity, while $\mu_2$ is the low-velocity branch, as $v_{r_2}$
is much smaller and becomes negligible at large radii.

\subsection{Soliton boundary conditions}
\label{sec:soliton-boundary}

We now derive the large-radius boundary condition required by a matching to the
soliton solution recalled in section~\ref{sec:large-radius-soliton}.
At large radii but within the soliton radius, $r_{\rm sg} \ll r \ll R_{\rm sol}$,
the scalar field is  in the weak-gravity regime dominated by the scalar cloud mass and  approaches
the core of the soliton solution (\ref{eq:soliton-rho-PhiI}).
As seen in the previous section, this also corresponds to the small-$\mu$ regime.
The comparison of the expression (\ref{eq:phi-small-mu}) with Eq.(\ref{eq:phi-soliton})
gives
\be
r_{\rm sg} \ll r \ll R_{\rm sol} : \;\;\;
\phi_0(r) = \frac{\sqrt{2\rho_{\rm sol}(0)}}{m}  ,
\;\;\; \beta \simeq 0 ,
\label{eq:phi0-rhos}
\ee
and Eq.(\ref{eq:omega0-alpha}), where $\alpha$ takes the value defined by the soliton
solution (\ref{eq:phi-soliton}).
Indeed, as the soliton solution (\ref{eq:phi-soliton}) corresponds to the hydrostatic
equilibrium with $\vec v=0$, the ``velocity'' $\beta'$ must become negligible at large
radii in order to match with the soliton.
The uniform angular frequency $\omega_0$ is also set by this large-radius boundary
condition.
As $|\alpha| \lesssim 10^{-5}$ from Eq.(\ref{eq:alpha-Phi-Rs}),
$\omega_0$ remains very close to $m$.

We can now check that this is consistent with Eq.(\ref{eq:beta-1}).
Combining Eqs.(\ref{eq:phi0-mu}) and (\ref{eq:phi0-rhos}) we obtain
\be
\mu = \frac{\rho}{\Lambda^4} = \frac{8}{3 |k_2|} \Phi_{\rm I} ,
\label{eq:mu-Phi-I}
\ee
where we introduced the repulsive potential defined in (\ref{eq:Phi-I-def}).
Equation (\ref{eq:beta-1}) with $\beta'=0$ gives, at leading order in $\Phi$ and $\mu$,
\be
\omega_0 = m \left( 1 + \Phi - 3 k_2 \mu/8 \right) = m (1+\Phi+\Phi_{\rm I})
\label{eq:omega0-Phi-k2}
\ee
where we used the series expansion (\ref{eq:Q-mu-series}) and in the last equality
we used Eq.(\ref{eq:mu-Phi-I}).
Then, using Eq.(\ref{eq:alpha-def}) we recover Eq.(\ref{eq:omega0-alpha}).
This shows that this large-radius asymptote is self-consistent.

Using $\Phi_{\rm I}=\alpha-\Phi$ from Eq.(\ref{eq:alpha-def}),
the comparison of Eq.(\ref{eq:mu-Phi-I}) with Eq.(\ref{eq:mu2-def})
shows that the matching to the soliton solution selects the branch $\mu_2(x)$
at large radii:
\be
\mu = \mu_2(x) \;\; \mbox{for} \;\;
r \gg r_{\rm sg} .
\label{eq:soliton-mu2}
\ee
This agrees with the fact that the branch $\mu_2(x)$ corresponds to the solution
with negligible radial velocity, as shown in Eq.(\ref{eq:vr-2}), which allows the matching
to the static soliton.

\subsection{Boundary condition at the horizon}

Close to the horizon the self-interactions cannot counteract the BH gravity and the scalar
field is in a free fall regime where $v_r \sim - \sqrt{h/f}$ \cite{Brax:2019npi},
with purely ingoing solutions and relativistic velocities.
Therefore, we must reach the high-velocity branch $\mu_1(x)$,
\be
\mu = \mu_1(x) \;\;\; \mbox{for} \;\;\;  r \simeq r_s/4 .
\label{eq:mu1-horizon}
\ee
This corresponds to the solution of Eq.(\ref{eq:Flux-Fs}), understood as an equation
for $\mu$, that is on the left side of the peak of $F(\mu)$.
From Eq.(\ref{eq:vr-sqrt}) we also recover $v_r \sim -  \sqrt{h/f}$ near the horizon
for this branch, as $f \to 0$ and the last square root becomes of order unity.

In the relativistic regime, $v_r$ can no longer be identified with a physical velocity
(e.g., the velocity of particles of mass $m$) as the Euler equation (\ref{eq:Euler-1})
no longer applies. This is why it can go to infinity, whereas the velocity of an infalling
particle measured by a distant observer would actually vanish at the horizon because
of a strong redshift. We shall see in section \ref{sec:Schwarzschild-behavior} that the divergence of $v_r$ at the horizon
is only an artefact and that the scalar field $\phi$ remains well behaved down to the
Schwarzschild radius.

\section{Quartic Lagrangian}
\label{sec:quartic-Lagrangian}

We consider in this section the case of the quartic scalar field Lagrangian
(\ref{eq:lambda4-k2}), where only an $X^2$ term is added to the standard kinetic term.
This provides a simple example where we cannot connect to the soliton at large radii
and there is no steady solution with a slow infall into the BH.

\subsection{Limited parameter range for the nonlinear oscillator}
\label{sec:limited-range-quartic}

In this quartic case, we have
\be
G(X) = X - \frac{3 |k_2|}{2} X^2 ,
\ee
and
\be
0 \leq y \leq \frac{1}{6|k_2|} : \;\;\; G^{-1}(y) = \frac{1-\sqrt{1-6|k_2|y}}{3|k_2|} .
\ee
Because $G(X)$ is only monotonically increasing on the finite interval
$0 \leq X \leq 1/(3|k_2|)$ and decreasing beyond, the inverse function $G^{-1}(y)$
is only defined over the finite range $0 \leq y \leq 1/(6|k_2|)$.
Following the analysis of section~\ref{sec:nonlinear-oscillator}, the differential equation
(\ref{eq:ck-ODE-def}) only admits a periodic oscillating solution defined for all real $u$
if $\mu < 1/(6|k_2|)$, when the quarter of period ${\bf Q}$ obtained in
Eq.(\ref{eq:Q-mu}) is well defined. It now reads as
\be
{\bf Q}(\mu) = \int_0^1 d\psi \sqrt{ \frac{3 |k_2| \mu}
{1-\sqrt{1+6 |k_2| \mu (\psi^2-1)}} } .
\ee
For larger values of $\mu$, the function ${\rm ck}(u)$ displays a singularity at
${\rm ck}=\sqrt{1-1/(6 |k_2| \mu)}$, with
$\frac{\partial{\rm ck}}{\partial u} = -1/\sqrt{3 |k_2| \mu}$
and $\frac{\partial^2{\rm ck}}{\partial u^2} = -\infty$.
Then, there is no regular periodic solution.
This implies that the oscillatory parameter $\mu$ is restricted to the finite range
\be
0 \leq \mu \leq \mu_{\max} \;\;\; \mbox{with} \;\;\; \mu_{\max} = \frac{1}{6 |k_2|} .
\label{eq:mu-upper-fixed}
\ee
It must also satisfy the upper bound (\ref{eq:mu+-def}), so that at each radius
$\mu(x)$ is restricted to the range
\be
0 \leq \mu \leq \min ( \mu_+(x) , \mu_{\rm max} ) .
\ee

\subsection{Flux $F(\mu,x)$ as a function of $\mu$}
\label{sec:flux-quartic}

\begin{figure}
\begin{center}
\epsfxsize=8.8 cm \epsfysize=6 cm {\epsfbox{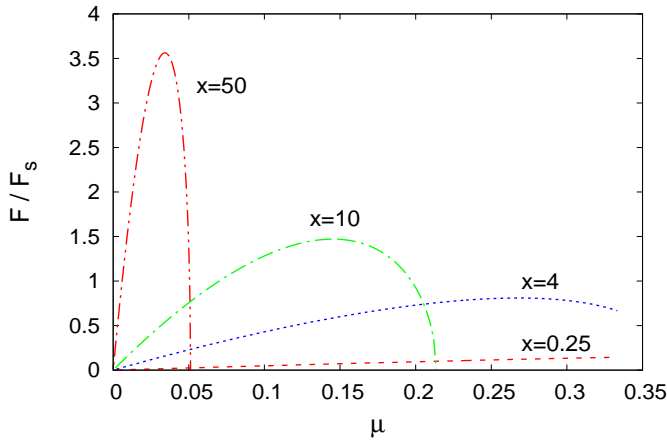}}
\end{center}
\caption{Normalized flux $F(\mu,x)/F_s$ as a function of the oscillatory parameter $\mu$,
for various values of the radial coordinate $x$, from Eq.(\ref{eq:Flux-Fs}) for the case
of the quartic Lagrangian (\ref{eq:k2-quartic}).}
\label{fig_flux-quartic}
\end{figure}

We show in Fig.~\ref{fig_flux-quartic} the flux $F(\mu,x)$ as a function of $\mu$
for several values of the radius $x$, for the case
\be
k_2 = - \frac{1}{2} , \;\;\; \mu_{\max} = \frac{1}{3} .
\label{eq:k2-quartic}
\ee
A change of $|k_2|$ can be absorbed in a change of $\mu$ at constant $|k_2|\mu$,
so this figure describes all Lagrangians with a quartic derivative self-interaction $X^2$.

The fixed upper bound (\ref{eq:mu-upper-fixed}), due to the fact that $G^{-1}(y)$ is not
defined for all positive $y$, makes an important change to the
generic behavior described above after Eq.(\ref{eq:mu+-def}) and seen to be universal
at large radii in section~\ref{sec:small-field-large-radius}.
At large radii we recover the single peak with a vanishing flux at both endpoints.
The peak moves to lower $\mu$ with an increasing height as the radius $x$ increases,
in agreement with section~\ref{sec:small-field-large-radius}.
However,  close to the Schwarzschild radius only part of the left side of the
peak can be reached as (\ref{eq:mu-upper-fixed}) prevents to access larger values of $\mu$ and in particular the peak itself.
This implies that a steady solution with constant flux $F$ can only exist for low
values of the flux, $F \leq F(\mu_{\max},1/4)$, below the maximal flux reached at
the Schwarzschild radius, $x=1/4$.
Moreover, starting from the horizon we can see that a regular solution can only follow
the left branch $\mu_1(x)$, on the left side of the peak.
This is inconsistent with the large-radius boundary condition (\ref{eq:soliton-mu2}),
which requires to be on the right branch $\mu_2$ at large radii.
Therefore, in contrast with the case of a quartic potential investigated
in \cite{Brax:2019npi}, there is no continuous and steady global solution that applies
down to the BH horizon while converging to the static soliton at large radii.
This means that, contrary to the quartic potential, the quartic derivative self-interaction
$-X^2$ is not able to support the scalar cloud against the BH gravity in the
relativistic regime.
Then, we expect the supermassive BH to ``eat'' the scalar cloud in a short time,
as compared with the age of the Universe.

\subsection{Constant flux solution}
\label{sec:continuous-solutions-quartic}

We now study in more details the continuous solutions defined by a constant flux $F$.
Because these solutions no longer connect to the soliton, the parameter
$\alpha$ introduced in Eq.(\ref{eq:omega0-alpha}), understood as the definition of the
common angular frequency $\omega_0$ in units of $m$, is no longer related to
the value $\alpha$ associated with the soliton in (\ref{eq:phi-soliton}).
In particular, we no longer have the relation (\ref{eq:alpha-def}) between the gravitational
and self-interaction potentials. Indeed, we no longer have hydrostatic equilibrium
at large radii.

Once we have chosen a value $F$ for the flux, which is below the maximum value
reached at the horizon $F_{\rm max}(x=1/4) \simeq 0.1 F_s$, the function $\mu(r)$
is set at all radii by the intersection of the left side of the peak of $F(\mu,x)$
with $F$, see Fig.~\ref{fig_flux-quartic}.
From $\mu(r)$ we obtain the amplitude $\phi_0(r)$ from Eq.(\ref{eq:phi0-mu}),
while $\beta'$ is given by Eq.(\ref{eq:beta-1}). This fully determines $\beta(r)$,
up to an irrelevant integration constant, and the scalar field (\ref{eq:phi-ck-def}) at radii
$r\ll r_{\rm sg}$, where the metric is dominated by the BH and given
by Eqs.(\ref{eq:f-def})-(\ref{eq:h-def}) or (\ref{eq:Phi-BH}).
At larger radii, the metric would be set by the scalar cloud self-gravity and we would solve
in a self-consistent manner the Poisson equation.

For $\mu \ll 1$ the scalar field reduces to the cosine (\ref{eq:phi-small-mu})
at lowest order, as ${\rm ck}(u,0) \simeq \cos(u)$, and the kinetic factor $X$ from
Eq.(\ref {eq:X-mu-ck2}) reads
\be
X = \mu \sin^2(\omega_0 t-\pi\beta/2) =
\frac{m^2 \phi_0^2}{2 \Lambda^4} \sin^2(\omega_0 t-\pi\beta/2) ,
\ee
where we used Eq.(\ref{eq:phi0-mu}).
In this small-$X$ regime we also have $K \simeq X$, $K' \simeq 1$, and
the scalar field energy density reads at lowest order
\be
r \gg r_s : \;\;\; \rho_\phi = \frac{m^2 \phi_0^2}{2} = \mu \Lambda^4 .
\label{eq:rho-phi-NR}
\ee
We recover the nonrelativistic regime, as in Eq.(\ref{eq:phi-soliton}) that
applied to the nonrelativistic soliton.
Using Eqs.(\ref{eq:mu1-def}) and (\ref{eq:Fs-def}) this also reads as
\be
r \gg r_s : \;\;\; \rho_\phi = \frac{|F|}{r^2\sqrt{2(\alpha-\Phi)}} = \frac{F}{r^2 v_r} ,
\label{eq:rho-F-NR}
\ee
where in the second equality we used Eq.(\ref{eq:vr-1}).
We recover the nonrelativistic definition of the flux, $F=r^2 \rho_\phi v_r$.

The gravitational potential becomes dominated by the scalar cloud gravity
when the mass $M_\phi$ associated with the scalar field within the radius $r$
is of the order of the black hole mass.
Writing $M_\phi \sim \rho_\phi r^3$ and $\alpha-\Phi \sim -\Phi \sim r_s/r$,
as $\Phi \sim \Phi_{\rm BH}$ up to this transition radius,
we obtain
\be
r_{\rm sg} \sim r_a \left( \frac{F}{F_s} \right)^{-2/3}
\left( \frac{r_a}{r_s} \right)^{1/3} .
\label{eq:rsg-too-large}
\ee
As $F/F_s<1$ from Fig.~\ref{fig_flux-quartic} and $r_a \gg r_s$, where $r_a$ is the
characteristic radius defined in Eq.(\ref{eq:r_a-rho_a}),
we obtain for the transition radius $r_{\rm sg} \gg r_a$.
This corresponds to a radius that is much larger than the soliton radius
$R_{\rm sol}=\pi r_a$, recalled below Eq.(\ref{eq:r_a-rho_a}).
Therefore, we conclude that for these continuous solutions all radii are
dominated by the BH gravity. This is because the high infall velocity,
$|v_r| \sim \sqrt{-\Phi}$, implies a much smaller density, $\rho_\phi = F/(r^2 v_r)$,
than for the second branch $\mu_2(r)$ that converges to the static soliton,
associated with the much smaller velocity (\ref{eq:vr-2}).
Thus, these solutions describe the latest stages of the infall of the
scalar cloud onto the central BH, when most of the scalar-field cloud
has already been ``eaten'' by the BH.

\subsection{Free-fall flux}
\label{sec:free-fall-flux-quartic}

Another interpretation of the result (\ref{eq:rsg-too-large}) can be obtained
from the flux expected in the case of free fall.
At large radii, $r \gg r_s$, the scalar field is  in the nonrelativistic regime and we expect
the scalar field cloud to fall into the central BH with the free-fall velocity
$v_r \sim - \sqrt{2{\cal G}M/r}$. Using $F = r^2\rho_\phi v_r$ in the nonrelativistic
regime and the definition of $F_s$ in Eq.(\ref{eq:Fs-def}), we obtain
\be
\frac{F}{F_s} \sim \left( \frac{M}{M_{\rm BH}} \right)^2 \left( \frac{r}{r_a} \right)^{-3}
\left( \frac{\rho}{\bar\rho_c} \right)^{-1/2} \frac{1}{H r_a} ,
\ee
where $\bar\rho_c$ is the cosmological critical density and $H$ the Hubble expansion rate.
For $r\sim r_a$, with $r_a \sim 20 {\rm kpc}$, $\rho \sim 10^6 \bar\rho_c$ and
$M \sim 10^3 M_{\rm BH}$, we obtain $F/F_s \sim 10^8$.
Such a large flux cannot be accomodated by the solutions shown in
Fig.~\ref{fig_flux-quartic}.
This explains why we found in Eq.(\ref{eq:rsg-too-large}) that the continuous
solution with the profile (\ref{eq:rho-F-NR}) can only describe at best the late stages
of the infall, after most of the scalar mass has disappeared into the BH and only
a small scalar mass remains, which can be transported with a small flux.
Therefore, the infall of the scalar cloud cannot be described by the oscillatory
solutions (\ref{eq:phi-ck-def}).

In any case, the fact that there are no regular solutions that satisfy both boundary
conditions, at the horizon and at the soliton core, shows that such scalar field models
cannot support a stable galactic-mass scalar cloud around a supermassive BH.
Therefore, they cannot provide realistic dark matter scenarios.
This shows the importance of checking the self-consistency of the system from the
galactic kpc scales down to the Schwarzschild radius and taking into account
the relativistic regime. Indeed, as recalled above, at large radii in the nonrelativistic
regime this model is equivalent to a quartic potential model and the derivative
self-interaction $-X^2$ builds an effective pressure that is able to support the
scalar cloud against gravity.
It happens that close to the Schwarzschild radius one enters the nonlinear regime,
where the $X^2$ term is no longer a small correction to the standard kinetic term,
and there the scalar field is no longer able to provide a self-consistent support against
the BH gravity.

At large $X$ the sign of the kinetic term also becomes negative, $K' <0$, which
typically signals the appearance of ghosts. Thus, such a theory is also problematic
at a more fundamental theoretical level. Hence, we do not consider  this
theory further.

\section{Conditions to stabilize the soliton}
\label{sec:power-law}

We have seen in the previous section that when the function $G^{-1}(y)$ is not
defined over all positive $y$, as happened for the quartic Lagrangian with $k_2<0$,
it may be impossible to obtain a steady state solution that satisfies both small
and large radii boundary conditions.
We investigate in this section the conditions
to obtain global solutions that can match the static soliton at large radii.
For simplicity, we focus on the large-$X$ behavior, where we assume that the kinetic
function $K(X)$ behaves as  a power-law (with $K'>0$),
\be
X \gg 1 : \;\;\; K(X) \simeq a X^{\nu} , \;\;\; a>0, \;\; \nu > 0 .
\label{eq:KX-nu-def}
\ee
This yields for the functions $G(X)$ and $G^{-1}(y)$
\be
G(X) = a (2\nu-1) X^{\nu} , \;\;\; G^{-1}(y) = \left[ \frac{y}{a(2\nu-1)} \right]^{1/\nu} ,
\ee
with the constraint
\be
\nu > \frac{1}{2} ,
\label{eq:nu-one-half}
\ee
so that $G(X)$ is monotonically increasing and $G^{-1}$ is well defined.
Then, the quarter of period ${\bf Q}$ of Eq.(\ref{eq:Q-mu}) reads
\be
{\bf Q} = \mu^{(\nu-1)/(2\nu)} [ a (2\nu-1) ]^{1/(2\nu)}
\frac{\sqrt{\pi} \Gamma[1-1/(2\nu)]}{2 \Gamma[3/2-1/(2\nu)]} .
\label{eq:Q-power-law}
\ee
As explained in section~\ref{sec:steady-quartic}, the flux $F(\mu,x)$ as a function
of $\mu$ generically shows a peak and vanishes at both endpoints of the range
(\ref{eq:mu+-def}).
The large-radius boundary condition selects the low-velocity branch (\ref{eq:mu2-def})
associated with the right side of the peak, while the small-radius boundary condition
selects the high-velocity branch (\ref{eq:mu1-def}) associated with the left side of
the peak.
To find a global solution that smoothly switches from the left to the right branch,
the height $F_{\rm peak}$ of the peak must increase at both large and small radii.
We have seen that this is always the case at large radii, in the nonrelativistic regime
(\ref{eq:Fm-small-mu}).
At small radii, this depends on the large-$X$ behavior of the kinetic function, and
we have seen in section~\ref{sec:quartic-Lagrangian} that for the quartic Lagrangian
this does not happen, as the peak is cut from the right by the additional
upper bound (\ref{eq:mu-upper-fixed}).
Then, although this is neither a sufficient nor a necessary condition, models that are likely
to show the desired behavior for $F(\mu,x)$ should show the full peak, with the vanishing
of the flux at both endpoints $0$ and $\mu_+$.
At the horizon the metric function $f(r)$ vanishes. From Eq.(\ref{eq:mu+-def})
this means that in order to reach $\mu_+(x)$, down to the horizon, the quarter of period
${\bf Q}$ must decrease down to zero for large $\mu$.
In the power-law case (\ref{eq:Q-power-law}) this gives the constraint
\be
\nu < 1 ,
\label{eq:nu-one-upper}
\ee
so that ${\bf Q}$ vanishes for $\mu\to\infty$.
From Eq.(\ref{eq:Cmu}) we obtain
\be
C_\mu = \frac{\nu}{3\nu-1} ,
\ee
and the flux (\ref{eq:Flux-Fs}) reads
\ba
&& \frac{F}{F_s} = x^2 h \frac{\nu [a (2\nu-1)]^{1/\nu} \Gamma[1-1/(2\nu)]^2}
{(3\nu-1) \pi \Gamma[3/2-1/(2\nu)]^2} \mu^{(2\nu-1)/\nu} \hspace{0.8cm}
\nonumber \\
&& \times \sqrt{1 - \frac{\pi \Gamma[3/2-1/(2\nu)]^2 f \mu^{(1-\nu)/\nu} }
{(1+\alpha)^2 [a (2\nu-1)]^{1/\nu} \Gamma[1-1/(2\nu)]^2} } . \hspace{0.6cm}
\ea
At the horizon $x$ and $h$ are finite while $f$ vanishes. We can see that
$\mu_+$ grows as $f^{-\nu/(1-\nu)}$ and the peak height grows as
$f^{-(2\nu-1)/(1-\nu)}$.
Therefore, we find indeed the required behavior for a global transonic solution,
as the peak of $F(\mu)$ is fully obtained at small radii with a height that grows
as we move closer to the horizon.

As we will explicitly note in section~\ref{sec:radial-profile}, the condition $1/2<\nu<1$
that we have obtained here is only suggestive. It is neither a necessary nor a sufficient
condition. For instance, kinetic functions $K(X)$ that show the asymptotic slope
$K(X) \sim X^{\nu}$ with $1/2<\nu<1$ but are badly behaved for intermediate
values of $X$, e.g. if they violate the condition (\ref{eq:Gp-positive}) at intermediate
$X$, cannot provide a realistic or physical model.
On the other hand, we shall see in section~\ref{sec:explicit}, on the example
(\ref{eq:KX-2o3}) of a well-behaved kinetic function $K(X)$, that $X$ actually remains
bounded down to the BH horizon, $0 \leq X \leq X_{\max}(x=1/4)$.
Then, the very large $X$ behavior of $K(X)$, at values that are not reached in practice,
is actually irrelevant.
It may however be probed by other configurations, e.g. in the early Universe.

\section{Explicit example}
\label{sec:explicit}

\subsection{Characteristic functions}
\label{sec:characteristic-functions}

\begin{figure}
\begin{center}
\epsfxsize=8.8 cm \epsfysize=6 cm {\epsfbox{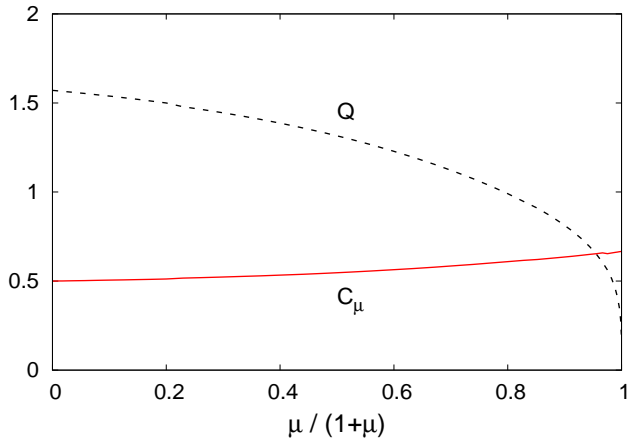}}
\end{center}
\caption{Quarter of period ${\bf Q}$, from Eq.(\ref{eq:Q-mu}), and
average $C_{\mu}$, from Eq.(\ref{eq:Cmu}), as functions of the
oscillatory parameter $\mu$. We use the ratio $\mu/(1+\mu)$ for the abscissa,
to cover the range $0 \leq \mu < \infty$.}
\label{fig_QC}
\end{figure}

\begin{figure}
\begin{center}
\epsfxsize=8.8 cm \epsfysize=6 cm {\epsfbox{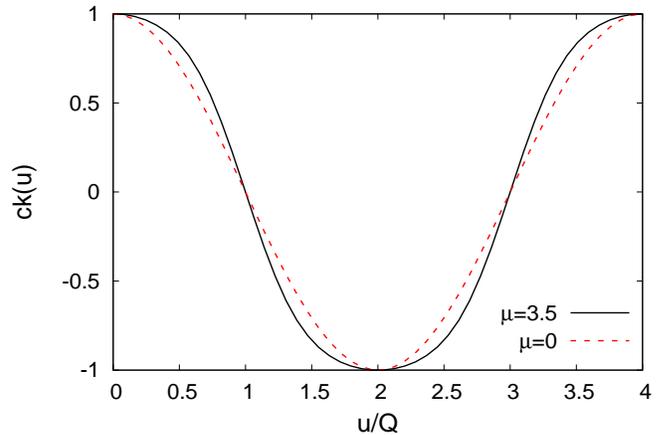}}
\end{center}
\caption{Nonlinear oscillatory function ${\rm ck}(u,\mu)$, for $\mu=3.5$
(solid line) and $\mu=0$ (dashed line). The abscissa is renormalized by the
quarter of period ${\bf Q}$.}
\label{fig_ck}
\end{figure}

To illustrate the results of the previous section we now consider the case
\be
K(X) = (1+3X/2)^{2/3} - 1 .
\label{eq:KX-2o3}
\ee
This corresponds to the exponent $\nu=2/3$, which falls in the range $1/2<\nu<1$
obtained in Eqs.(\ref{eq:nu-one-half}) and (\ref{eq:nu-one-upper}).
This also gives the quadratic coefficient $k_2=-1/2$, as for the quartic model
(\ref{eq:k2-quartic}),
\be
\nu = 2/3 , \;\;\; k_2 = -1/2 .
\ee
The function $G(X)$ reads as
\be
G(X) = \frac{1}{3} (1+3X/2)^{2/3} - \frac{4}{3} (1+3X/2)^{-1/3} + 1 ,
\ee
and the inverse function $G^{-1}(y)$ as
\ba
G^{-1}(y) & = & \frac{2}{3} \biggl [  \frac{ (-1 + y + (2+\sqrt{5-3y+3y^2-y^3})^{2/3} )^3 }
{2+\sqrt{5-3y+3y^2-y^3}} \nonumber \\
&& -1 \biggl ] .
\ea
At large $y$ this gives the power-law asymptote
\be
y \to \infty : \;\;\; G^{-1}(y) = 2 \sqrt{3} y^{3/2} + \dots .
\ee
The nonlinear differential equation (\ref{eq:ck-E}) now admits regular periodic solutions
${\rm ck}(u,\mu)$ for all positive $\mu$.
Then, at any radius $x$ the oscillatory parameter $\mu$ is only bounded by
$\mu_+(x)$ from Eq.(\ref{eq:mu+-def}).

We show in Fig.~\ref{fig_QC} the quarter of period ${\bf Q}$ and the average
$C_{\mu}$ as functions of $\mu$.
In agreement with the analysis of section~\ref{sec:power-law}, ${\bf Q}$ goes to zero
as $\mu$ goes to infinity, while $C_{\mu}$ remains finite and does not vary much
over the full range $0\leq \mu < \infty$.
More precisely, we have the asymptotic behaviors
\be
\mu \to \infty : \;\;\; {\bf Q} = \frac{\sqrt{2\pi} \Gamma[5/4]}{3^{1/4} \Gamma[3/4]}
\mu^{-1/4} , \;\;\; C_{\mu} = \frac{2}{3} .
\ee
We shall find in section~\ref{sec:critical} that $\mu$ grows at smaller radii and reaches
at the horizon the value $\mu_s \simeq 3.5$ of Eq.(\ref{eq:mu-s-def}) below.
This corresponds to ${\bf Q}_s \simeq 1$ and $C_{\mu_s} \simeq 0.6$.
Therefore, the oscillatory parameter $\mu$ reaches the mildly nonlinear regime
at the Schwarzschild radius. There, all higher-order terms of the kinetic function
$K(X)$ are relevant, as is obvious from the fact that the physics associated with
the kinetic function (\ref{eq:KX-2o3}) will be quite different from the one associated
with the quartic case analyzed in section~\ref{sec:quartic-Lagrangian}
(a quasistatic soliton can now be supported around the supermassive BH).

We display in Fig.~\ref{fig_ck} the oscillatory function ${\rm ck}(u,\mu)$ for
$\mu=0$, where ${\rm ck}(u,0)=\cos(u)$, and for $\mu=\mu_s\simeq 3.5$.
We only show the first period, $0 \leq u \leq 4 {\bf Q}$, and we renormalize the
abscissa by ${\bf Q}$. We can see that although ${\bf Q}$ has decreased from
${\bf Q}(0)=\pi/2$ to ${\bf Q}_s \simeq 1$, the shape of the function ${\rm ck}(u)$
remains close to the cosine once we renormalize the period.

\subsection{Critical solution}
\label{sec:critical}

\begin{figure}
\begin{center}
\epsfxsize=8.8 cm \epsfysize=6 cm {\epsfbox{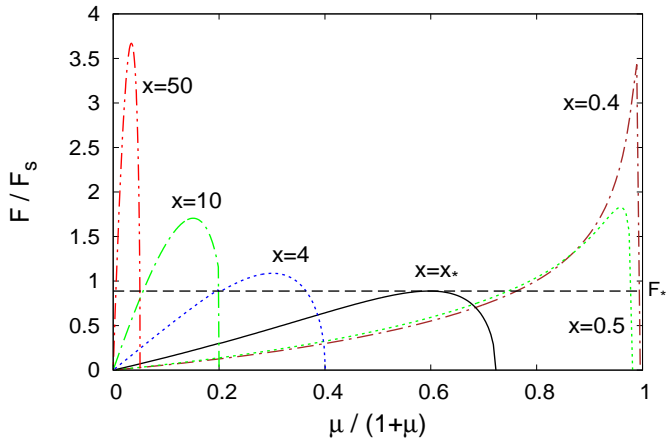}}
\end{center}
\caption{Normalized flux $F(\mu,x)/F_s$ as a function of the oscillatory parameter $\mu$,
for various values of the radial coordinate $x$, from Eq.(\ref{eq:Flux-Fs}) for the
kinetic function (\ref{eq:KX-2o3}). We use the ratio $\mu/(1+\mu)$ for the abscissa,
so that large values of $\mu$ fit into the figure.
The horizontal dashed line is the value $F_\star$ of Eq.(\ref{eq:x-star}), defined
as the minimum over all radii, $1/4 \leq x < \infty$, of the height of the peak.
It is reached at radius $x_\star \simeq 1.4$.}
\label{fig_flux}
\end{figure}

\begin{figure}
\begin{center}
\epsfxsize=8.8 cm \epsfysize=6 cm {\epsfbox{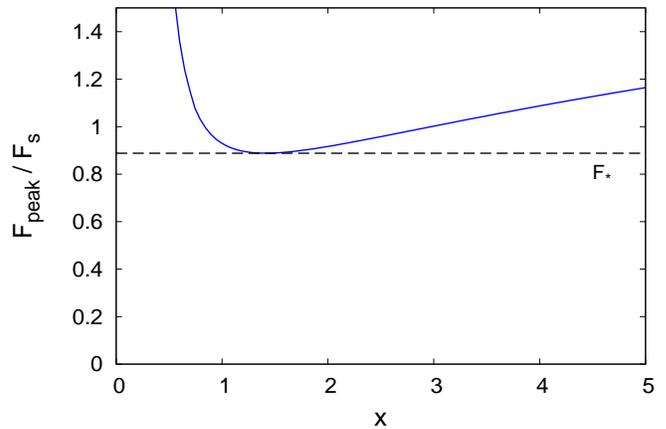}}
\end{center}
\caption{Peak value $F_{\rm peak}(x)/F_s$ as a function of the radial coordinate $x$.
The horizontal dashed line is the minimum value $F_\star \simeq 0.9$, reached
at $x_\star \simeq 1.4$.}
\label{fig_Fmax}
\end{figure}

We show in Fig.~\ref{fig_flux} the flux $F(\mu,x)$ as a function of $\mu$
for several values of the radius $x$.
Close to the horizon the peak moves to large values of $\mu$, as
$\mu_+ \to \infty$ for $x \to 1/4$. To display the curves from $x=50$ to
$x=0.4$ on the same plot we use the abscissa $\mu/(1+\mu)$ in Fig.~\ref{fig_flux}.
In agreement with the analysis of the previous section, we now find that $F(\mu)$
shows a full peak, with a vanishing flux at both endpoints, at any radius $x$.
Moreover, the peak height increases for both large and small radii, with a minimum
$|F_c|=F_\star |F_s|$ at the intermediate radius $x_\star$,
\be
\frac{F_c}{F_s} = F_\star \;\;\; \mbox{with}  \;\;\; x_\star \simeq 1.4, \;\;\;
F_\star \simeq 0.9 .
\label{eq:x-star}
\ee
We show the curve $F_{\rm peak}(x)/F_s$ of the peak height, as a function of the radius $x$,
in Fig.~\ref{fig_Fmax}. This clearly shows the increase of the peak height at small and large
radii and the minimum at $x_\star$.

\begin{figure}
\begin{center}
\epsfxsize=8.8 cm \epsfysize=6 cm {\epsfbox{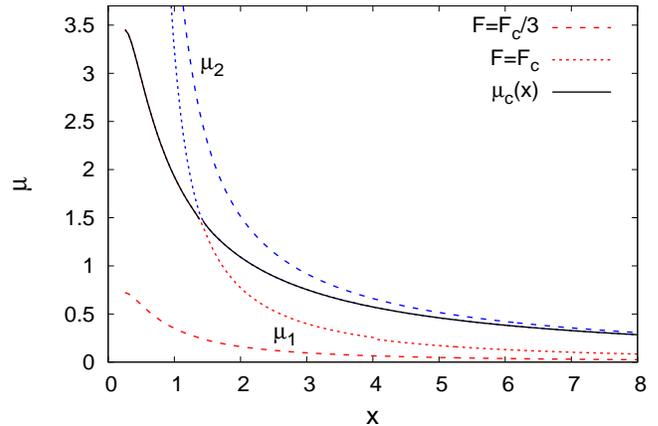}}
\end{center}
\caption{Oscillatory parameters $\mu_1(x)$ and $\mu_2(x)$ for a constant flux $F_c/3$
(dashed lines) and $F_c$ (dotted lines). The critical curve $\mu_c(x)$ (solid line) is equal
to $\mu_1$ for $x<x_\star$ and to $\mu_2$ for $x>x_\star$, with $F=F_c$.}
\label{fig_mu1_mu2}
\end{figure}

A global solution is obtained provided the constant flux $|F|$ is smaller than
the critical value $|F_c|$, so that at any radius $x$ above the horizon there is at least
one solution $\mu(x)$ to Eq.(\ref{eq:Flux-Fs}), understood as an implicit equation for $\mu$.
For $|F| < |F_c|$ there are two solutions $\mu_1 < \mu_2$ at any radius and we recover
the high- and low-velocity branches discussed below (\ref{eq:mu+-def}) and analyzed
in the limit of large radii in section~\ref{sec:small-mu}.
We have seen in (\ref{eq:mu1-horizon}) that at small radii, near the horizon,
we must follow the left branch $\mu_1(x)$, associated with large infall velocities.
On the other hand, we have seen in (\ref{eq:soliton-mu2}) that at large radii
we must follow the right branch $\mu_2$, associated with small infall velocities,
to converge to the static soliton.
These two boundary conditions select the critical value $F_c$ as the only physical
value for the flux, which allows us to obtain a regular solution $\mu_c(x)$ that
smoothly connects the two branches at radius $x_\star$, where they meet.

We show the two branches $\{\mu_1(x),\mu_2(x)\}$ in Fig.~\ref{fig_mu1_mu2}
for the critical flux $F_c$ and for a lower flux $F_c/3$.
In the case $F_c/3$, these two branches remain well separated at all radii
and we cannot switch from one side of the peak to the other.
Only for the critical value $F_c$ we can switch from $\mu_1$ to $\mu_2$
in a continuous manner, as shown in the figure.
This is similar to the selection of the unique transonic solution in the hydrodynamical case
\cite{Michel:1972}.
At large radii, in the weak gravity regime, $\mu_c(x)=\mu_2(x)$ is given by
Eq.(\ref{eq:mu2-def}). Close to the horizon $\mu_c(x)=\mu_1(x)$ remains finite
and our numerical computation shown in Fig.~\ref{fig_mu1_mu2} gives
\be
r \to r_s/4 : \;\;\; \mu_c(x) \to \mu_s \;\;\; \mbox{with} \;\;\; \mu_s \simeq 3.5 .
\label{eq:mu-s-def}
\ee
The fact that $\mu$ is of order unity near the horizon shows that the scalar
field is in the nonlinear regime, where all nonlinear terms in the kinetic function
$K(X)$ are relevant.

\subsection{Radial profile}
\label{sec:radial-profile}

\begin{figure}
\begin{center}
\epsfxsize=8.8 cm \epsfysize=6 cm {\epsfbox{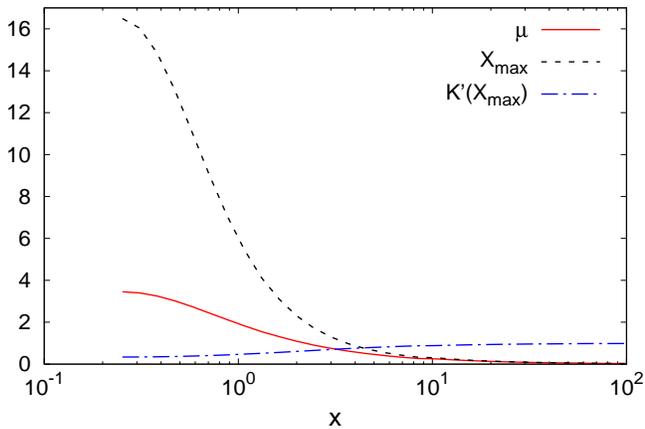}}
\end{center}
\caption{Radial profile of the oscillatory parameter $\mu(x)$,
of the upper bound $X_{\rm max}(x)$ of Eq.(\ref{eq:Xmax-r}),
and of the first derivative $K'[X_{\rm max}(x)]$.}
\label{fig_muX}
\end{figure}

Thus, the scalar field $\phi(r,t)$ depends on both radius and time as determined by
the solution (\ref{eq:phi-ck-def}), where the parameter $\mu$ of the
nonlinear oscillator ${\rm ck}(u,\mu)$ follows the critical solution $\mu_c(x)$ displayed
in Fig.~\ref{fig_mu1_mu2} above.
At each radius, the kinetic argument $X$ oscillates with time, as seen in Eq.(\ref {eq:X-mu-ck2}).
Using Eq.(\ref{eq:dck-du-Gm1}), this also reads as $X = G^{-1}[\mu(1-{\rm ck}^2)]$,
so that $X$ oscillates in the range
\be
0 \leq X(r,t) \leq  X_{\rm max}(r) \;\;\; \mbox{with} \;\;\; X_{\rm max} = G^{-1}(\mu) .
\label{eq:Xmax-r}
\ee
We show again in Fig.~\ref{fig_muX} the radial profile of the oscillatory parameter
$\mu(x)$, given by the critical curve $\mu_c(x)$ of Fig.~\ref{fig_mu1_mu2}, as well
as the upper value $X_{\rm max}(x)$.
We can see that $X_{\rm max} \simeq 16.5$ and $K'(X_{\rm max}) \simeq 0.33$ at the horizon.
Although we reach the nonlinear regime, $K'$ is still of order unity.
Because $X$ remains finite at the horizon, the function $K(X)$ can deviate from
the expression (\ref{eq:KX-2o3}) for $X>X_{\max}(r_s/4)$ without changing our results.
This implies that the function $K(X)$ is not required to show the power-law behavior
$K \sim X^{2/3}$ at infinity and a broader class of kinetic functions are able to support
the scalar field soliton around the supermassive BH.

\subsection{Behavior at the Schwarzschild radius}
\label{sec:Schwarzschild-behavior}

\subsubsection{Isotropic coordinates}
\label{sec:Schwarzschild-isotropic}

As for the case of the quartic potential \cite{Brax:2019npi},
the effective velocity $v_r=\pi\beta'/(2m)$ diverges
 at the Schwarzschild radius because of the metric factor $1/f$.
Thus, from Eq.(\ref{eq:vr-sqrt}) we obtain close to the Schwarzschild radius,
where $f \to 0$,
\ba
r \to r_s/4 : && \frac{\pi \beta'}{2m} \sim - (1+\alpha) \sqrt{\frac{h}{f}}
\sim - \frac{(1+\alpha) 8 r_s}{4 r - r_s} , \nonumber \\
&& \beta \sim - \frac{(1+\alpha) 4 m r_s}{\pi} \ln \left( \frac{4 r - r_s}{4 r_s} \right) .
\hspace{0.5cm}
\label{eq:beta-rs}
\ea
However, this divergence is only an artefact, due to the use of the isotropic coordinates
(\ref{eq:ds2-Schwarzschild-isotropic}).

\subsubsection{Eddington coordinates}
\label{sec:Schwarzschild-Eddington}

As in \cite{Brax:2019npi}, to check that the scalar field remains well behaved down to the
horizon it is convenient to introduce the Eddington coordinates $(\tilde{t},\tilde{r})$,
where $\tilde r$ is the standard Schwarzschild radial coordinate of Eq.(\ref{eq:tilde-r-def})
and $\tilde{t}$ is the Eddington time, defined by \cite{Blau-2017}
\be
\tilde{t} = t + r_s \ln \left| \frac{\tilde r}{r_s} - 1 \right| .
\label{eq:Eddington-time}
\ee
This gives the metric
\ba
ds^2 & = & - \left(1-\frac{r_s}{\tilde r} \right) d\tilde{t}^2
+ 2 \frac{r_s}{\tilde r} d\tilde{t} d\tilde{r}
+ \left(1+\frac{r_s}{\tilde r}\right) d\tilde{r}^2 \nonumber \\
&& + \tilde r^2 d\vec\Omega^2 ,
\label{eq:ds2-Eddington}
\ea
which is regular over all $\tilde{r} > 0$.
These coordinates $(\tilde{t},\tilde{r})$ are directly related to the
Eddington-Finkelstein coordinates \cite{Blau-2017}.

Substituting the result (\ref{eq:beta-rs}) into Eq.(\ref{eq:phi-ck-def}) gives
\be
\tilde{r} \to r_s : \;\;\; \phi = \phi_s \,
{\rm ck} \left[ \frac{2 {\bf Q}_s}{\pi} (1+\alpha) m ( \tilde{t}+\tilde{r} )
, \mu_s \right] ,
\label{eq:phi-rs}
\ee
where the parameter $\mu_s$ at the Schwarzschild radius was obtained in
Eq.(\ref{eq:mu-s-def}) and the amplitude $\phi_s$ is given by Eq.(\ref{eq:phi0-mu})
in terms of $\mu_s$.
As for the free scalar and for the case of a quartic potential \cite{Brax:2019npi},
the scalar field is well defined at the
horizon and we recover a purely ingoing solution with unit velocity.
Nevertheless, the derivative self-interactions remain relevant down to the horizon
as (\ref{eq:phi-rs}) differs from the cosine (i.e. harmonic) expression
of the free case. We obtain a nonlinear radial wave,
with higher-order harmonics given by the expansion (\ref{eq:ck-Fourier}).

We can now come back to the definition of the velocity. In the large radius limit,
we have identified $v_r=\pi\beta'/(2m)$ of Eq.(\ref{eq:vr-sqrt}) with the velocity obtained
in the fluid picture of the nonrelativistic dark matter through the Euler equation (\ref{eq:Euler-1})
and Eq.(\ref{eq:v-s-def}).
In fact, Eq.(\ref{eq:beta-1}) allows us to go beyond this large-radius regime.
Indeed, we can identify this relation with the relativistic dispersion relation
of a particle of mass $m$ and momentum $p^{\mu}$,
$g_{\mu\nu} p^\mu p^\nu = - m^2$, with
\be
p^0 = \frac{2 {\bf Q} \omega_0}{\pi f} , \ \ p^r = \frac{{\bf Q} \beta'}{h} .
\ee
Then the speed can be identified as
\be
c^r = \frac{p^r}{p^0} = \frac{f}{h} \frac{\pi \beta'}{2 \omega_0} ,
\ee
which coincides with $v_r$ in the large-radius limit and for $\alpha \ll 1$.
Close to the BH horizon, we have seen that Eq.(\ref{eq:phi0-mu}) gives
\be
r \to r_s/4 : \;\;\; \beta' \simeq - \frac{2\omega_0}{\pi} \sqrt{ \frac{h}{f} }
\ee
as $f\to 0$ while ${\bf Q}$ remains finite, see also Eq.(\ref{eq:beta-rs}).
This yields
\be
r \to r_s/4 : \;\;\; c^r \simeq - \sqrt{ \frac{f}{h} } .
\ee
If we use the Schwarzschild radial coordinate $\tilde r$ instead of the isotropic
radial coordinate $r$, we have from Eq.(\ref{eq:tilde-r-def})
$d\tilde{r}/dr = \sqrt{fh}$ and we obtain
\be
c^{\tilde r} = \sqrt{ \frac{f^3}{h} } \frac{\pi \beta'}{2 \omega_0} ,
\ee
and
\be
\tilde{r} \to r_s : \;\;\; c^{\tilde r} \simeq - f .
\ee
We find that both velocities vanish at the horizon. We recover the well known result
that the velocity of infalling matter measured by a distant observer (at rest at infinity,
with a proper time given by $t$) vanishes as the body approaches the horizon.
On the other, the proper time of a particle that is falling from infinity at rest is
$d\tau = f dt$ \cite{Shapiro:1983du}. Thus, we recover $d\tilde r/d\tau = -1$ at the horizon
following the infalling matter. The dynamics become highly relativistic as we approach the
horizon.

\subsection{Density profile}
\label{sec:Density-profile-lambda4}

\begin{figure}
\begin{center}
\epsfxsize=8.8 cm \epsfysize=6 cm {\epsfbox{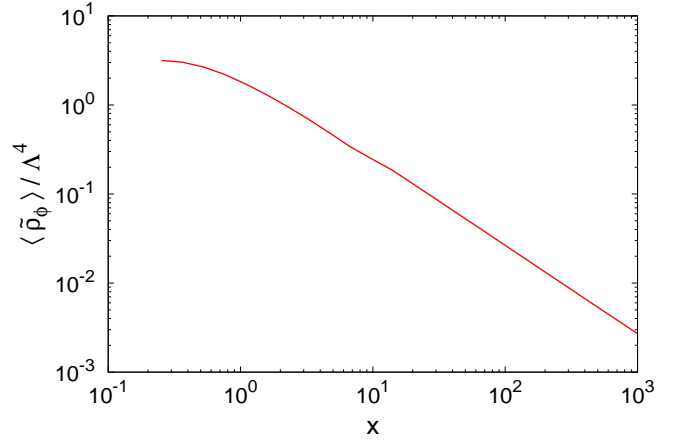}}
\end{center}
\caption{Scalar-field energy density computed in the Eddington metric, from
the Schwarzschild radius up to $10^4 r_s$, where the metric potentials are still dominated
by the central BH. We plot the average $\langle \tilde\rho_\phi \rangle$ in the fast
oscillations with time.}
\label{fig_rho}
\end{figure}

For spherically symmetric configurations, the density defined
by the time-time component of the energy-momentum tensor in the coordinates
$(\tilde{t},\tilde{r})$ reads
\ba
\tilde\rho_\phi \equiv - \tilde{T}^0_0 & = & (2-f) K' \left(
\frac{\partial \phi}{\partial\tilde t} \right)^2 + (f-1) K'
\frac{\partial \phi}{\partial\tilde t} \frac{\partial \phi}{\partial\tilde r}
\nonumber \\
&& - \Lambda^4 K + \frac{m^2}{2} \phi^2 ,
\label{eq:rho-Eddington}
\ea
and the partial derivatives are related by
\be
\frac{\partial \phi}{\partial\tilde t} = \frac{\partial \phi}{\partial t} , \;\;\;
\frac{\partial \phi}{\partial\tilde r} = \frac{\partial \phi}{\partial r}
\frac{1}{\sqrt{f h}} + \frac{\partial \phi}{\partial t} \left( 1 - \frac{1}{f} \right) .
\label{eq:derivative-Eddington}
\ee
For the solution (\ref{eq:phi-ck-def}), with Eqs.(\ref{eq:dphi-dt-ck}) and
(\ref{eq:dphi-dr-ck}), this gives
\ba
\frac{\tilde\rho_\phi}{\Lambda^4} & = & \mu \, {\rm ck}^2 - K
+ K' \left( \frac{\partial{\rm ck}}{\partial u} \right)^2 2\mu
\left[ \frac{2{\bf Q}}{\pi} (1+\alpha) \right]^2 \nonumber \\
&& \times \biggl \lbrace  \frac{1}{f}
+ \frac{f-1}{f} \sqrt{ 1 - \frac{\pi^2 f}{(1+\alpha)^2 4 {\bf Q}^2} } \biggl \rbrace .
\label{eq:rho-ck}
\ea
This energy density remains finite at the Schwarzschild radius.
Neglecting $\alpha \ll 1$, we obtain
\be
\tilde{r} = r_s , \;\; r = \frac{r_s}{4} : \;\;\; \langle \tilde\rho_\phi \rangle \simeq
3.2 \, \Lambda^4 \simeq 0.6 \, \rho_a ,
\label{eq:T00-Lambda4-rs}
\ee
where $\langle \dots \rangle$ denotes the average over the fast oscillations
over time, as in Eq.(\ref{eq:C-mu-def})-(\ref{eq:Cmu}).
Contrary to the case of the free scalar, where the flux $F$ and the density
$\tilde\rho_\phi$ can take any value, for the self-interacting scalar field
$F$ and $\tilde\rho_\phi$ are uniquely determined (because the system becomes
nonlinear). As could be expected, the density (\ref{eq:T00-Lambda4-rs})
is set by the characteristic density $\rho_a$ defined in Eq.(\ref{eq:Phi-I-def}),
which measures the strength of the self-interactions.
The unboundedness of the free case is recovered by the fact that
$\langle \tilde\rho_\phi \rangle \to \infty$ when $\rho_a \to \infty$,
which corresponds to vanishing self-interactions, $k_2 \to 0$.

In the weak-gravity regime, dominated by the BH, $\mu(x)$ follows the low-velocity
branch $\mu_2(x)$ of Eq.(\ref{eq:mu2-def}). This gives
\be
r_s \ll r \ll r_{\rm sg} : \;\;\; \mu \simeq - \frac{16 \Phi}{3} \simeq \frac{8 r_s}{3 r} ,
\ee
where we used Eq.(\ref{eq:Phi-BH}).  In this regime $\mu \ll 1$ and at leading order
the density (\ref{eq:rho-ck}) gives
\be
r_s \ll r \ll r_{\rm sg} : \;\;\;
\frac{\langle \tilde\rho_\phi \rangle}{\Lambda^4} \simeq \mu \simeq \frac{8 r_s}{3 r}
\propto r^{-1}
\label{eq:profile-rho}
\ee
while the velocity (\ref{eq:vr-2}) reads
\be
r_s \ll r \ll r_{\rm sg} : \;\;\; v_r \simeq - \frac{3 F_\star}{4} \frac{r_s}{r}
\propto r^{-1} .
\ee
As for the case of a quartic potential, the density decreases with radius
as $1/r$, more slowly than the $r^{-3/2}$ falloff obtained for the free scalar
\cite{Brax:2019npi}.
This is because the velocity decreases faster, as $1/r$ instead of $1/\sqrt{r}$,
as the self-interactions give rise to an effective pressure that stabilizes the scalar
cloud and enables the convergence to the static soliton solution at large radii.
We show in Fig.~\ref{fig_rho} the scalar field profile obtained from
Eq.(\ref{eq:rho-ck}), averaged over the fast oscillations.
It clearly displays the $1/r$ profile (\ref{eq:profile-rho}) at large radii
and the finite value (\ref{eq:T00-Lambda4-rs}) at the horizon.

\subsection{Lifetime of the scalar-field soliton}
\label{sec:lifetime}

At large radii, in the weak gravity regime dominated by the scalar cloud self-gravity,
the velocity (\ref{eq:vr-2}) reads
\be
r_{\rm sg} \ll r \lesssim R_{\rm sol} : \;\;\;
v_r \simeq - \frac{3 F_\star}{8} \frac{\rho_a}{\rho_{\rm sol}} \frac{r_s^2}{r^2} ,
\ee
where we used Eqs.(\ref{eq:alpha-def}) and (\ref{eq:Phi-I-def}).
This coincides with the result obtained in \cite{Brax:2019npi} for the case
of a quartic potential. Indeed, as recalled in (\ref{eq:lambda4-k2}), in the
nonrelativistic small-field regime the derivative self-interaction is equivalent
to a potential self-interaction.
Again, at radii of the order of the soliton radius $R_{\rm sol} = \pi r_a$,
this gives the typical radial velocity $v_r$ and evolution timescale $t_c$
\be
v_r(r_a) \sim - \frac{\rho_a}{\rho_{\rm sol}} \frac{r_s^2}{r_a^2}  , \;\;\;
t_c \equiv \frac{r_a}{|v_r|} \sim r_a \frac{\rho_{\rm sol}}{\rho_a} \frac{r_a^2}{r_s^2} .
\ee
This also reads
\be
t_c \sim t_H \left( \frac{\bar\rho_c}{\rho_a} \right)^{5/2}
\frac{\rho_{\rm sol}}{\bar\rho_c} \left( \frac{R_H}{r_s} \right)^2 ,
\ee
where $t_H=1/H$ and $R_H=1/H$ are the Hubble time and Hubble radius,
and $\bar\rho_c=3H^2/(8\pi{\cal G})$ is the cosmological critical density.
At redshift $z=0$ this gives
\be
t_c \sim 10^3 \, t_H \frac{\rho_{\rm sol}}{\bar\rho_c}
\left( \frac{\rho_a}{1\,{\rm eV}^4} \right)^{-5/2}
\left( \frac{M}{10^8 M_\odot} \right)^{-2} .
\ee
For the soliton to give rise to a significant departure from the CDM profiles
on galactic scales, we must have a radius of about $20$ kpc, which gives
$\rho_a \sim 1 \, {\rm eV}^4$ \cite{Brax:2019fzb}.
Larger characteristic densities lead to smaller soliton radii.
We typically have $\rho_{\rm sol}/\bar\rho_c \sim 10^5$ for the DM overdensity
in the soliton core. Therefore, we find that $t_c \gg t_H$
and the DM solitonic cores can easily survive until today, despite the
infall of the inner layers into the central supermassive BH.

Again, astrophysical stellar mass BHs cannot eat a significant fraction
of the galactic DM soliton.
Indeed, for $N$ BHs of unit solar mass, the typical timescale
for the soliton depletion reads
\be
t_N \sim 10^{19} \, \frac{t_H}{N} \frac{\rho_{\rm sol}}{\bar\rho_c}
\left( \frac{\rho_a}{1\,{\rm eV}^4} \right)^{-5/2} .
\ee
Since we typically have $N < 10^{11}$, as only a fraction of the galactic baryonic
mass can be made of stellar BHs, we obtain $t_N \gg 10^8 t_H$ and
the soliton mass loss is negligible.

\section{Renormalization of the action}
\label{sec:renormalization}

We have seen in the previous sections that for nonlinear kinetic functions that
satisfy conditions such as (\ref{eq:nu-one-half}) and (\ref{eq:nu-one-upper})
the scalar field with the Lagrangian (\ref{eq:action-S-def}) displays well behaved solutions
from the Newtonian to the relativistic regimes, i.e. from the small-field and weak-gravity
to the large-field and strong-gravity regimes.
In this section, we check that quantum corrections remain small and do not invalidate
the previous analysis, based on the classical equations of motion.

\subsection{Weak-gravity regime}
\label{sec:weak-gravity-quantum}

In the spirit of background quantization, we decompose the scalar field
in the classical background $\bar\phi$, which is a solution of the classical
equations of motion, and the quantum fluctuations $\hat\phi$,
\be
\phi= \bar \phi +\hat \phi .
\label{eq:phi-back-fluc}
\ee
The kinetic argument also reads  $X=\bar X + \hat X$, with
\be
\hat X = - \frac{1}{\Lambda^4} g^{\mu\nu} \partial_\mu\bar\phi \partial_\nu\hat\phi
- \frac{1}{2\Lambda^4} g^{\mu\nu} \partial_\mu\hat\phi \partial_\nu\hat\phi ,
\label{eq:hat-X-def}
\ee
while the Lagrangian can be expanded as ${\cal L} = \bar{\cal L} + \hat{\cal L}$, with
\be
\hat {\cal L} = \Lambda^4 \left( \bar K' \hat X + \frac{\bar K''}{2} \hat X^2 + \dots \right)
- \frac{m^2}{2} (2 \bar \phi \hat\phi + \hat\phi^2 ) .
\label{eq:dL-def}
\ee

We first consider the weak-gravity regime, far from the BH, where the background geometry
is well described by the Minkowski spacetime and the background scalar field
is scale independent,
\be
\bar\phi = \bar\phi(t) , \;\;\; \frac{d\bar\phi}{dt}  \sim m \bar\phi .
\ee
This corresponds for instance to the core of the static soliton solution
(\ref{eq:phi-soliton}), where ${\vec v}=0$ and the phase $s$ only depends on time.
Expanding the Lagrangian to second order in the perturbation, we get
from Eq.(\ref{eq:dL-def}) the second-order variation
\be
\hat {\cal L}^{(2)}= \frac{\bar K'+2 \bar X \bar K''}{2}
\left( \frac{\partial \hat\phi}{\partial t} \right)^2 - \frac{\bar K'}{2} (\nabla \hat\phi)^2
- \frac{m^2}{2} \hat\phi^2 .
\label{eq:L2-Minkowski}
\ee
This can be the basis of a well-defined perturbation theory,
without ghosts nor small-scale instabilities in the linear regime, when we have:
\be
\bar K' > 0 , \;\;\; \bar K'+2 \bar X \bar K'' > 0 .
\label{eq:Kp-Ks-ghost}
\ee
We implicitly assumed $K' \geq 0$ throughout this article.
The second condition, $K'+2 X K'' \geq 0$, coincide with the condition
(\ref{eq:Gp-positive}) that was required to build the solution (\ref{eq:phi-ck-def})
from a well-defined nonlinear oscillatory function ${\rm ck}(u)$, as described in
section~\ref{sec:nonlinear-oscillator}.

It is convenient to normalize the field $\hat\phi$ as
\be
\hat\phi= \frac{\varphi}{\sqrt{\bar K' + 2 \bar X \bar K''}} .
\ee
After one integration by parts, this gives
\be
\hat {\cal L}^{(2)}= \frac{1}{2} \left( \frac{\partial \varphi}{\partial t} \right)^2
- \frac{c_s^2}{2} (\nabla \varphi)^2 - \frac{\bar m^2}{2} \varphi^2 ,
\ee
where the speed of sound $c_s$ is defined by
\be
c_s^2= \frac{\bar K'}{\bar K'+ 2 \bar X \bar K''} > 0 ,
\ee
and the effective mass $\bar m$
\be
\bar m^2= \frac{m^2}{\bar K' + 2 \bar X \bar K''}  - \frac{
\frac{d^2}{dt^2} \left[ \sqrt{\bar K' + 2 \bar X \bar K''} \right] }
{\sqrt{\bar K' + 2 \bar X \bar K''}} .
\label{eq:m-bar-def}
\ee
At low $\bar X$, that is, for small amplitude of the scalar field and
$\bar\rho_\phi \ll \Lambda^4$, we have $\bar K \simeq \bar X \ll 1$ and
\be
\bar X \ll 1 : \;\;\; c_s^2 \simeq 1, \;\;\; \bar m \simeq m .
\label{eq:cs-low-X}
\ee
At large $\bar X$, for a power-law behavior $K(X) \sim X^\nu$ as in Eq.(\ref{eq:KX-nu-def}),
we have
\be
\bar X \gg 1 : \;\;\; c_s^2 \simeq \frac{1}{2\nu-1} > 0 , \;\;\;
\bar m \sim \frac{m}{\sqrt{\bar K'}} \gg m > 0 ,
\label{eq:cs-large-X}
\ee
provided the exponent $\nu$ verifies
\be
1/2 < \nu < 1 .
\label{eq:nu-half--one}
\ee
For $\nu > 1$ we still have $c_s^2>0$ but $\bar K'$ becomes large and the squared mass
becomes negative, as it is dominated by the second term in Eq.(\ref{eq:m-bar-def})
which scales as $ - (\nu-1)^2 m^2$ for $d\bar X/dt \sim \pm m \bar X$.
The bounds (\ref{eq:nu-half--one}) coincide with the bounds (\ref{eq:nu-one-half})
and (\ref{eq:nu-one-upper}) that were required at the classical level to obtain well-behaved
global solutions, from the strong-gravity to the weak-gravity regimes.
Therefore, they are satisfied by realistic models, such as Eq.(\ref{eq:KX-2o3}).

We are interested in quantum phenomena in the ultraviolet (UV). In the infrared (IR)
there are no divergences thanks to the scalar mass $m$, which is much larger than
cosmological scales. Hence we neglect the time variation
of $c_s$ and $\bar m$. From Eqs.(\ref{eq:cs-low-X})-(\ref{eq:cs-large-X})
$c_s$ is always of order unity, therefore we take $c_s \sim 1$ and omit factors $c_s$
in the order-of-magnitude estimates below.
Then, the propagator for the quantum field $\varphi$ behaves like
\be
G_\varphi(\omega,\vec p \, )= \frac{1}{-\omega^2 + {\vec p^{\,2}} + \bar m^2} =
\frac{1}{p^2+\bar m^2} .
\label{eq:propagator-Minkowski}
\ee
Let us now consider the interaction terms. They spring from expressions like
$\Lambda^4 \bar K^{(n)} \frac{\hat X^n}{n!}$.
In the following, we omit numerical factors and focus on the scalings with $\bar X$.
Then, we write (\ref{eq:hat-X-def}) as
\be
\hat X \sim \bar X^{1/2} \left( \frac{\partial\hat\phi}{\Lambda^2} \right) +
\left( \frac{\partial\hat\phi}{\Lambda^2} \right)^2 ,
\ee
and we obtain for the cubic and higher-order terms of the Lagrangian,
\be
\hat{\cal L}^{(n\geq 3)} = \Lambda^4 \sum_{n=3}^{\infty} \hat{c}_n
\left( \frac{\partial\hat\phi}{\Lambda^2} \right)^n ,
\label{eq:Ln-vertex-hat-phi}
\ee
with
\be
\hat{c}_n = \sum_{m=[n/2]_+}^n \bar K^{(m)} \bar X^{m-n/2} ,
\ee
where $[n/2]_+$ is the smallest integer that is greater than or equal to $n/2$.
In terms of the rescaled field $\varphi$, this gives
\be
\hat{\cal L}^{(n\geq 3)} = \Lambda^4 \sum_{n=3}^{\infty} c_n
\left( \frac{\partial\varphi}{\Lambda^2} \right)^n ,
\label{eq:Ln-vertex-varphi}
\ee
with
\be
c_n = ( \bar K'+2 \bar X \bar K'' )^{-n/2} \sum_{m=[n/2]_+}^n \bar K^{(m)} \bar X^{m-n/2} .
\label{eq:cn-def}
\ee
In the weak-field regime, $\bar X \ll 1$, the sum (\ref{eq:cn-def}) is dominated by
the first term $m=[n/2]_+$ and even-order terms are of the order of unity while odd-order
terms are of the order of $\bar X^{1/2}$,
\be
\bar X \ll 1 : \;\;\; c_{2n} \sim 1 .
\label{eq:c-2n-low-X}
\ee
On the other hand, in the strong-field regime, $\bar X \gg 1$, using
$\bar K^{(n)} \sim \bar K / \bar X^n$ for the power law $K \propto X^{\nu}$,
all terms in the sum (\ref{eq:cn-def}) contribute and we obtain
\be
\bar X \gg 1 : \;\;\; c_{n} \sim \bar K^{1-n/2} \ll 1 .
\label{eq:c-n-large-X}
\ee

From the propagator (\ref{eq:propagator-Minkowski}) and the vertices
(\ref{eq:Ln-vertex-varphi}), a typical $L$-loop vacuum Feynman diagram contributing
to the corrections to the classical action reads
\be
I_L= \int \prod_{\ell=1}^L d^4 p_\ell \prod_{n=1}^N \frac{1}{p^2+\bar m^2}
\prod_{v=1}^V  \Lambda^4 c_{m_v} \prod_{s=1}^{m_v} \frac{p_s}{\Lambda^2}
\ee
where there are $N$ propagators corresponding to $N$ lines in the diagram and
$V$ vertices, each with a degree $m_v$.
Rescaling momenta by $\bar m$, using the Euler identity  $V-N=1-L$ and
$\sum_{v=1}^V m_v = 2 N$, we obtain
\be
I_L = \Lambda^4 \left( \frac{\bar m}{\Lambda} \right)^{4L} \left( \prod_{v=1}^V c_{m_v} \right)
\tilde I_L ,
\ee
where the integral $\tilde I_L$ is dimensionless and does not depend on $\bar m$,
$\Lambda$ nor $\bar K$.
It is divergent and needs to be regularized, for instance using dimensional regularization.
The infinite part appears as poles in $1/(d-4)$, where $d$ is the dimension of spacetime.
Removing these infinities requires to introduce counterterms in the bare action.
This leaves finite corrections to the classical action that scale as
\be
L \geq 1 : \;\;\; \delta {\cal L}^{(L)} \sim  \Lambda^4 \left( \frac{\bar m}{\Lambda} \right)^{4 L}
\left( \prod_{v=1}^V c_{m_v} \right) .
\ee
Notice that this expression depends on the background field via $\bar m$,
the coefficients $c_{m_v}$ and the sound speed $c_s$ (which we omit in the expressions).
There are two types of corrections. The first ones involve the second term in the
expression (\ref{eq:m-bar-def}) of the effective mass $\bar m$ and depend
on higher derivatives $\partial \bar X$ and $\partial^2 \bar X$,
i.e. second and third derivatives of $\bar \phi$.
If they were the only types of corrections, we would retrieve the usual nonrenormalization
theorem of $K(X)$ theories.
The second ones involve the first term only in the expression (\ref{eq:m-bar-def})
of $\bar m$, $m^2/(\bar K' + 2 \bar X \bar K'')$, as well as factors of $c_s$ and $c_{m_v}$.
These corrections depend on $\bar X$ and provide corrections to the classical Lagrangian
$K(\bar X)$. Hence in the models considered here the classical Lagrangian is renormalized,
because the bare mass is nonzero.
Nevertheless, these quantum corrections can remain negligible, as we now investigate.

First, in the weak-field regime, $\bar X \ll 1$, we obtain from
(\ref{eq:cs-low-X}) and (\ref{eq:c-2n-low-X})
\be
L \geq 1 : \;\;\; \delta {\cal L}^{(L)} \sim  \Lambda^4 \left( \frac{m}{\Lambda} \right)^{4 L} .
\label{eq:loop-L-X1}
\ee
Therefore, higher loop corrections are under control and become increasingly small
at higher orders provided
\be
m \ll \Lambda : \;\;\;  \delta {\cal L}^{(L)} \ll \delta {\cal L}^{(1)}  \;\; \mbox{for} \;\; L \geq 2 .
\label{eq:crit1}
\ee
We must now compare the leading one-loop term to the classical action, ${\cal L}^{(0)}$,
\be
{\cal L}^{(0)} \sim \Lambda^4 \bar X - \frac{m^2}{2} \bar \phi^2 \sim \bar\rho_\phi
\gtrsim \bar\rho_0 ,
\ee
where $\bar\rho_0$ is the mean density of the Universe at redshift $z=0$.
This gives
\be
m^4 \ll \bar\rho_0 : \;\;\; \delta {\cal L}^{(1)} \ll {\cal L}^{(0)} ,
\ee
which reads
\be
m \ll 10^{-3} \ {\rm eV} .
\label{eq:crit2}
\ee

Second, in the strong-field regime, $\bar X \gg 1$, we obtain from (\ref{eq:cs-large-X})
and (\ref{eq:c-n-large-X})
\be
L \geq 1 : \;\;\; \delta {\cal L}^{(L)} \sim  \Lambda^4 \bar K
\left( \frac{m}{\Lambda} \right)^{4 L} ( \bar K \bar X^{-2/3} )^{-3L} .
\ee
Therefore, higher loop corrections do not blow up provided $\bar K \bar X^{-2/3}$ does not
go to zero at large $\bar X$. For the power-law behavior (\ref{eq:KX-nu-def}) this gives
the two conditions
\be
m \ll \Lambda \;\; \mbox{and} \;\; \nu \geq \frac{2}{3} : \;\;\;
\delta {\cal L}^{(L)} \ll \delta {\cal L}^{(1)}  \;\; \mbox{for} \;\; L \geq 2 .
\label{eq:crit1-large-X}
\ee
The classical action is now of the order of ${\cal L}^{(0)} \sim \Lambda^4 \bar K$.
As $\nu \geq 2/3$ ensures $\bar K \bar X^{-2/3} \gtrsim 1$ and we have
$m \ll \Lambda$, the conditions (\ref{eq:crit1-large-X}) also give
$\delta {\cal L}^{(1)} \ll {\cal L}^{(0)}$.

Therefore, the quantum corrections remain small for any scalar-field background, in the
weak gravity regime, provided we have the three conditions
\be
m \ll \Lambda , \;\;\; m \ll 10^{-3} \ {\rm eV} \;\; \mbox{and} \;\;
\nu \geq \frac{2}{3} \; \mbox{at large} \; X .
\label{eq:quantum-conditions}
\ee
The condition $\nu \geq 2/3$ is satisfied by the model (\ref{eq:KX-2o3}).
However, even at the BH horizon we have $\mu \lesssim 3.5$ from
Eq.(\ref{eq:mu-s-def}), i.e. $\bar X \sim 1$.
Therefore, even in this high-density region we do not probe the regime $\bar X \gg 1$
and we do not really need to satisfy the asymptotic condition $\nu \geq 2/3$ to keep the
quantum corrections negligible.
From Eq.(\ref{eq:lambda4-k2}), we note that for $k_2 \sim 1$ the quantum stability
implies $\lambda_4 \ll 1$.

\subsection{Schwarzschild background metric}
\label{sec:strong-gravity-quantum}

\begin{figure}
\begin{center}
\epsfxsize=8.8 cm \epsfysize=6 cm {\epsfbox{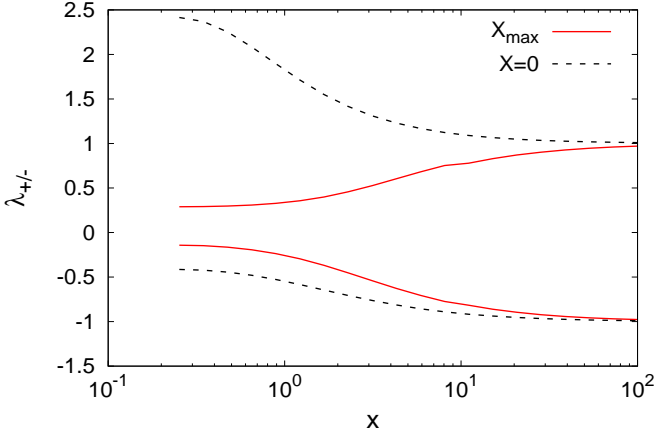}}
\end{center}
\caption{Radial profile of the eigenvalues $\lambda_+ > 0 > \lambda_-$
of the kinetic matrix ${\cal K}_{ij}$ that appears in Eq.(\ref{eq:d2L-Kij}).
We show the results obtained for $\bar X=X_{\rm max}$ (solid lines) and
for $\bar X=0$ (dashed lines).}
\label{fig_lambda}
\end{figure}

We now go beyond the Minkowski spacetime and consider the Schwarzschild metric
(\ref{eq:ds2-Schwarzschild-isotropic}) for the background.
This is valid from the large-radius weak-gravity regime (\ref{eq:Phi-BH}), which
is already covered by the analysis of the previous section, down to the BH horizon
in the strong-gravity regime.
The background scalar field $\bar\phi(r,t)$ now depends on both radius and time,
following the solution (\ref{eq:phi-ck-def}) described by the nonlinear oscillator
${\rm ck}(u,\mu)$.
At each radius, the background kinetic argument $\bar X$ oscillates
with time in the range given in (\ref{eq:Xmax-r}).

As in section~\ref{sec:Schwarzschild-Eddington}, we now work with the background
Eddington metric (\ref{eq:ds2-Eddington}), to be able to study the scalar field
down to the BH horizon.
We again obtain the Lagrangian of the fluctuations $\hat\phi$ from
Eq.(\ref{eq:dL-def}), paying attention to the fact that $\bar\phi$ now depends on both
time and radius.
In particular, using Eqs.(\ref{eq:derivative-Eddington}), (\ref{eq:dphi-dt-ck})
and (\ref{eq:dphi-dr-ck}), the derivatives of the background solution (\ref{eq:phi-ck-def})
with respect to the Eddington coordinates read
\be
\frac{\partial\bar\phi}{\partial\tilde t} = \phi_0 \omega \frac{\partial{\rm ck}}{\partial u} ,
\ee
and
\be
\frac{\partial\bar\phi}{\partial\tilde r} = \phi_0
\left[ \omega \left(1 - \frac{1}{f} \right) - \frac{{\bf Q}\beta'}{\sqrt{f h}} \right]
\frac{\partial{\rm ck}}{\partial u} .
\ee
Notice that $\frac{\partial\bar\phi}{\partial\tilde r} \neq \frac{\partial\bar\phi}{\partial\tilde t}$
because there are additional $\tilde r$-dependent terms to the one explicitly
written in Eq.(\ref{eq:phi-rs}), such as the radial dependence of the oscillatory parameter
$\mu(r)$.
Then, the Lagrangian of the fluctuations $\hat\phi$ reads at second order
\ba
\hat {\cal L}^{(2)} & = & \frac{1}{2} \left[ {\cal K}_{00}
\left( \frac{\partial \hat\phi}{\partial \tilde t} \right)^2
+ 2 {\cal K}_{01} \frac{\partial \hat\phi}{\partial \tilde t} \frac{\partial \hat\phi}{\partial \tilde r}
+  {\cal K}_{11} \left( \frac{\partial \hat\phi}{\partial \tilde r} \right)^2 \right]
\nonumber \\
&& - \frac{\bar K'}{2 \tilde r^2} (\partial_\Omega \hat \phi)^2 - \frac{m^2}{2} \hat\phi^2 ,
\label{eq:d2L-Kij}
\ea
where $\partial_\Omega \hat \phi$ is the angular derivative, with respect to the longitudinal
and azimuthal angles, and the coefficients ${\cal K}_{ij}$ are given by
\ba
{\cal K}_{00} & = & (2-f) \bar K' + \frac{2 \bar X \bar K''}{f^2 h m^2}
\biggl [ (f-1)^2 f ({\bf Q} \beta')^2   \nonumber \\
&& + 2 \sqrt{f h} (1-f) \omega {\bf Q} \beta' + h \omega^2 \biggl ] ,
\ea
\be
{\cal K}_{01} = (f-1) \bar K' + \frac{2 \bar X \bar K''}{f h m^2} \left[ (1-f) f ({\bf Q} \beta')^2
+ \sqrt{f h} \omega {\bf Q} \beta' \right] ,
\ee
\be
{\cal K}_{11} = -f \bar K' + \frac{2 \bar X \bar K'' f ({\bf Q} \beta')^2}{h m^2} .
\ee
We recover the scale-independent Minkowski case (\ref{eq:L2-Minkowski})
for $f=h=1$, $\omega=m$ and $\beta'=0$.
For $f \neq 1$ or $\beta' \neq 0$ we now have a mixing of the time and radial derivatives
in the kinetic term.
Using Eq.(\ref{eq:omega2-betap}), we find that the determinant of the kinetic matrix
${\cal K}_{ij}$, with ${\cal K}_{10}={\cal K}_{01}$, takes the simple form
\be
\det ( {\cal K}_{ij} ) = - \bar K' ( \bar K' + 2 \bar X \bar K'' ) < 0 .
\label{eq:det-K}
\ee
Here we assumed that the constraints (\ref{eq:Kp-Ks-ghost}) are already satisfied
by the kinetic function $K(X)$.
Remarkably, it coincides with the determinant obtained in the Minkowski case
(\ref{eq:L2-Minkowski}) as it does not depend on the metric potentials $f$ and $h$
nor on $\beta'$, but only on the properties of the kinetic function $K(X)$.
Its negative sign implies that the quadratic form governing the kinetic terms in the
$(\tilde t,\tilde r)$ plane has two opposite-sign eigenvalues $\lambda_+ > 0 > \lambda_-$.
This always preserves the signature $(+,-)$ and guarantees the absence of ghost
and gradient instability.
Moreover, as $\det ( {\cal K}_{ij} )$ does not vanish the two branches
$\lambda_+(x)$ and $\lambda_-(x)$ are well separated and do not make contact.
Therefore, the positive eigenvalue is connected to the eigenvector
$\frac{\partial\bar\phi}{\partial\tilde t}$ at large radii, while the negative eigenvalue is connected
to the eigenvector $\frac{\partial\bar\phi}{\partial\tilde r}$.
Close to the horizon, the eigenvectors are a linear combination of
$\frac{\partial\bar\phi}{\partial\tilde t}$ and $\frac{\partial\bar\phi}{\partial\tilde r}$.
However, one could define new time and radial coordinates $\hat t$ and $\hat r$, from
linear combinations of $\tilde t$ and $\tilde r$, so that $\hat t$ and $\hat r$ converge to
$\tilde t$ and $\tilde r$ at large radii and the kinetic term takes the diagonal form
$\frac{1}{2} [ \lambda_+ ( \frac{\partial\bar\phi}{\partial\hat t} )^2
+ \lambda_- ( \frac{\partial\bar\phi}{\partial\hat r} )^2 ]$.
We can check this behavior in Fig.~\ref{fig_lambda}, where we consider the two boundaries
$0$ and $X_{\max}$ of Eq.(\ref{eq:Xmax-r}) of the range spanned by the oscillating
background $\bar X(r,t)$.
Therefore, the second-order Lagrangian $\hat {\cal L}^{(2)}$ can be the basis of
a well defined quantum perturbation theory.

Because of the nondiagonal kinetic matrix ${\cal K}_{ij}$, the propagator will
be different from the  Minkowski rescaled propagator (\ref{eq:propagator-Minkowski}).
However, by going for instance to the diagonal coordinates $\{ \hat t, \hat r \}$
and using the fact that $\lambda_{\pm}$ remain of order unity, the scalings that we obtained
in the previous section~\ref{sec:weak-gravity-quantum} in the regime $\bar X \sim 1$
and $\bar K ' \sim 1$ remain valid.
In particular, as in Eq.(\ref{eq:propagator-Minkowski}) each propagator brings a factor
$1/m^2$ and as in Eq.(\ref{eq:Ln-vertex-hat-phi}) vertices
take the form $\Lambda^4 (\partial\hat\phi/\Lambda^2)^n$, with coefficients
$c_n$ of the order of unity.
Therefore, the power counting of loop diagrams is not altered and we recover
Eq.(\ref{eq:loop-L-X1}), while the classical Lagrangian is now of order
${\cal L}^{(0)} \sim \Lambda^4 \bar K \sim \Lambda^4$ as $\bar K \sim 1$.
Then, quantum loop corrections are small provided $m \ll \Lambda$,
\be
m \ll \Lambda : \;\;\; \delta {\cal L}^{(L)} \ll {\cal L}^{(0)} .
\ee
This condition was already required in (\ref{eq:crit1}) for the Minkowski background,
therefore the classical analysis developed in previous sections remains valid down
to the horizon.

Finally notice that the strong-coupling scale $\Lambda$ is not the
cutoff of the quantum theory. Indeed, nothing prevents one from using classical backgrounds
where $\rho_\phi \sim \Lambda^4$ as long as the quantum corrections are under control,
i.e. as long as the conditions (\ref{eq:quantum-conditions}) are satisfied.

\section{Conclusion}
\label{sec:conclusion}

We have shown in a previous article that a scalar field with a nonstandard kinetic term
can play the role of dark matter in the late Universe and build static solitonic profiles
in galaxies, with a flat core. In this weak-gravity and weak-field regime, the
first quartic correction $-(\partial\phi)^4$ to the kinetic term is the dominant subleading
correction. It provides an effective pressure that balances the self-gravity of the scalar
cloud and gives rise to a static equilibrium.

In this paper, we have investigated the impact of a supermassive BH at the center of
galaxies on this scenario.
To this order, following the spirit of our previous work for the case of a quartic subleading
potential $\phi^4$, which also gives rise to an effective pressure in the weak-gravity
and weak-field regime, we have obtained the explicit solution of the scalar-field equation
of motion in the large scalar mass limit.
For a $\phi^4$ correction to the potential, the nonlinearity transformed the usual harmonic
wave solution of the standard Klein-Gordon equation into a nonlinear wave described by
the Jacobi elliptic function ${\rm cn}(u,k)$.
In a similar fashion, we show how arbitrary kinetic functions $K(X)$ lead
to associated nonlinear oscillatory functions ${\rm ck}(u,\mu)$, which extend the
harmonic cosine $\cos(u)$ and the Jacobi elliptic function ${\rm cn}(u,k)$.
They correspond to second-order ordinary differential equations with nonlinear derivative terms,
which reduce to the harmonic oscillator at linear order.
In the large-mass limit, the scalar field shows fast oscillations with time, with an
angular frequency $\omega$ of order $m$, with an amplitude and a phase that
show a slow dependence on the distance from the central BH.

Contrary to the case of a quartic potential $\phi^4$, we find that the quartic derivative
self-interaction $-(\partial\phi)^4$ is not able to support the scalar cloud down to the
BH horizon, in the relativistic regime. A continuous solution can only describe
the late stage of the infall, when most of the scalar field energy density has already fallen
into the BH. Therefore, such models cannot provide realistic dark matter scenarios.
This shows the importance of going beyond the weak-gravity large-radius analysis and
of studying the self-consistency of the system down to the horizon, in the relativistic regime
and strong-gravity regime.

We discussed the generic conditions to obtain a well-behaved global solution.
We obtain the usual conditions $K'>0$ and $K'+2X K'' > 0$, which are typically associated
with the stability of perturbations in k-essence models, i.e. the absence of ghost and
the positivity of the squared  speed of sound.
We also note that in order to have a global solution, which satisfies the boundary condition
at the horizon and converges to the static soliton at large radii, $K(X)$ must typically grow
as a power law $X^{\nu}$ with $1/2 < \nu < 1$ at large $X$.

We have presented a detailed analysis of a simple well-behaved example,
$K(X)=(1+3X/2)^{2/3}-1$. There, in a fashion similar to both the hydrodynamical case
and the quartic potential case, a unique global solution exists. It is associated with
a critical value of the flux that allows the solution to match the boundary conditions
at both small and large radii.
This solution is well defined down to the horizon, once we use appropriate coordinates
such as the Eddington time. The amplitude of the scalar field $\phi$ and of the kinetic
argument $X$ remain finite at the Schwarzschild radius.
It leads to a slow infall of the scalar cloud into the BH, as the radial velocity
grows from a negligible value at large radii, in the quasistatic solitonic regime,
to relativistic values that follow the free infall at the horizon.
The infall timescale is much greater than the age of the Universe, hence these models
can provide realistic scenarios for the dark matter galactic halos.

Finally, we investigated the importance of quantum corrections to the classical action.
We obtained the conditions for these quantum corrections to remain negligible
in the configurations that we study in this paper, both in the weak-gravity regime,
well described by the Minkowski background metric, and the strong-gravity regime,
described by the Schwarzschild metric.
We recover the usual constraints $K'>0$ and $K'+2X K'' > 0$ for a well behaved
setup, and we find that when $m \ll \Lambda$ and $m \ll 10^{-3} \, {\rm eV}$
the quantum corrections remain small.
This holds both for small and large scalar field values, which can probe the strongly
coupled regime (there, in addition $K(X)$ must typically grow as $X^{\nu}$ with
$2/3 \leq \nu  < 1$).

Thus, we find that scalar fields with  nonstandard kinetic terms can provide
realistic models of dark matter, building solitonic cores at galactic centers that are
stable over the age of the Universe. Moreover, the quantum corrections remain
well under control. The conditions on the kinetic function $K(X)$ are mostly the
standard constraints, $K'>0$ and $K'+2X K'' > 0$, with the addition of a more
subtle condition associated with the existence of a global solution, which roughly
corresponds to $K(X) \sim X^{\nu}$ with $2/3 \leq \nu  < 1$ over $1 \lesssim X \lesssim 50$.
The scalar mass must obey $m \ll 10^{-3} \, {\rm eV}$ while the strong-coupling scale
must verify $\Lambda \gg m$.

\section{Acknowledgements}

  This work is
supported in part by the EU Horizon 2020 research and innovation
programme under the Marie-Sklodowska grant No. 690575. This article is
based upon work related to the COST Action CA15117 (CANTATA) supported
by COST (European Cooperation in Science and Technology).
The work by JARC is partially supported by the MINECO (Spain)
project FIS2016-78859-P(AEI/FEDER, UE).

\vspace{-.3cm}

\bibliography{ref2}


\end{document}